\documentclass[10pt,journal,compsoc]{IEEEtran}

\usepackage[colorlinks,urlcolor=blue,linkcolor=blue,citecolor=blue]{hyperref}
\usepackage{multibib}
\newcites{primary}{Primary Studies}

\usepackage{tabularray, fontawesome}
\usepackage[normalem]{ulem}
\usepackage{rotating}
\usepackage{color, float,soul}
\usepackage{tabularray, fontawesome}
\usepackage{array}

\usepackage{subcaption} 
\usepackage{enumitem, nicefrac, booktabs}

\usepackage{textcomp}
\usepackage{stfloats}
\usepackage{url}
\usepackage{verbatim}
\usepackage{graphicx}
\usepackage{cite}

\usepackage{array}
\usepackage{multirow}
\usepackage{bigstrut}
\usepackage{units}
\usepackage{flushend}
\usepackage{hyperref}
\usepackage{colortbl}
\usepackage{makecell}
\usepackage{mathptm}
\usepackage{tcolorbox}
\usepackage{fancyvrb} 
\usepackage{fvextra} 
\usepackage{microtype}
\usepackage{amsmath, amssymb}
\usepackage{tabularx}

\usepackage{enumitem}

\newcolumntype{P}[1]{>{\RaggedRight\hspace{0pt}}p{#1}}
\definecolor{tableGray}{RGB}{243, 244, 245}
\definecolor{borderGray}{RGB}{229, 230, 233}

\newtcolorbox{boxA}{
    colback = tableGray, 
    boxrule = 0pt  
}

\def\blackbar#1#2{
   {\color{black}\rule{#1mm}{4pt}}  #2}

\def\multicolorbar#1#2#3#4#5#6{
   {\color{black!20}\rule{#1mm}{6pt}}%
   {\color{black!40}\rule{#2mm}{6pt}}%
   {\color{black!60}\rule{#3mm}{6pt}}%
   {\color{black!75}\rule{#4mm}{6pt}}%
   {\color{black!100}\rule{#5mm}{6pt} #6}%
}

\newtcolorbox{highlightbox}{
	colback=gray!20, 
	colframe=black,   
	boxsep=2pt,       
	arc=0pt,          
	boxrule=1pt       
}

\title{Walking the Tightrope of LLMs for Software Development: A Practitioners' Perspective}

\author{Samuel Ferino, Rashina Hoda, John Grundy, Christoph Treude
\IEEEcompsocitemizethanks{\IEEEcompsocthanksitem S. Ferino, R. Hoda, and J. Grundy are with the Faculty of Information Technology,
Monash University, Melbourne, Australia. E-mail: \{samuel.demouraferino, rashina.hoda, john.grundy\}@monash.edu \protect\\
\IEEEcompsocthanksitem C. Treude is with the School of Computing and Information Systems, Singapore Management University, Singapore. E-mail: ctreude@smu.sg\protect\\
}
}

\begin{document}
\IEEEtitleabstractindextext{%
\begin{abstract}--\textbf{Background:} Large Language Models emerged with the potential of provoking a revolution in software development (e.g., automating processes, workforce transformation). Although studies have started to investigate the perceived impact of LLMs for software development, there is a need for empirical studies to comprehend how to balance forward and backward effects of using LLMs. \textbf{Objective:} We investigated how  LLMs impact software development and how to manage the impact from a software developer's perspective. \textbf{Method:} We conducted 22 interviews with software practitioners across 3 rounds of data collection and analysis, between October (2024) and September (2025). We employed Socio-Technical Grounded Theory for \textit{Data Analysis} (STGT4DA) to rigorously analyse interview participants' responses. \textbf{Results:} We identified the benefits (e.g., maintain developer flow, improve developer mental models, and foster entrepreneurship) and challenges (e.g., damage to developers' reputation) of using LLMs at individual, team, organisation, and society levels; as well as actionable guidances into how mitigate these  challenges. \textbf{Conclusion:} Critically, we present the trade-offs that software practitioners, teams, and organisations face in working with LLMs. Our findings are particularly useful for software team leaders and IT managers to assess the viability of LLMs within their specific context.

\end{abstract}

}
\maketitle

\IEEEpeerreviewmaketitle

\section{Introduction} \label{sec:introduction}

 \begin{quote}
     \textit{``There is nothing permanent except change."} -- \textsl{Heraclitus}
 \end{quote}

This quote by the Greek philosopher Heraclitus highlights how the world is continuously changing. Large Language Models (LLMs) are the contemporary catalyst for a revolution in the Information Technology sector \cite{ozkaya:2023}, starting from the release of LLM tools like ChatGPT and GitHub Copilot for the general public between 2022 and 2023 \cite{teubner:2023, dwivedi:2023}. LLM-powered code generators and assistants like GitHub Copilot, for instance, fostered the emergence of a new potential pillar for software development: \textsl{AI pair programming} \cite{dakhel:2023, zhou:2024}. LLMs can support a variety of software development tasks, such as code generation and information retrieval \cite{hou:2024, wang:2024, di:2025}. Enterprise LLM adoption reports from McKinsey \cite{mayer:2025} and DORA \cite{dora:2024, dora:2025} call attention to growing interest from companies in examining the potential of LLMs for software development. There is also a growing shift from traditional Q\&A online communities like \textit{Stack Overflow} towards LLM tools as the first source of support \cite{ozkaya:2023, burtch:2024}. For instance, the decline in networking traffic of \textit{Stack Overflow} may be attributed to developers adopting LLMs \cite{kabir:2024}.

Many investigations (e.g., \cite{ziegler:2024, cui:2024, ebert:2023, nam:2024}) highlight the potential benefits associated with software practitioners adopting LLMs. For instance, Cui et al. \cite{cui:2024} found that software developers using GitHub Copilot achieved an increase of 26.08\% in the number of weekly completed tasks when conducting an experiment with almost five thousand software developers from companies including Microsoft and Accenture. On the other hand, many studies (e.g., \cite{kuhail:2024, krauss:2025, lee:2025, chen:2025}) present the downsides of using LLMs. For instance, Lee et al. \cite{lee:2025} observed from surveying 319 knowledge workers that LLMs can potentially affect software developers' critical thinking skills. 

Although there is an emerging amount of studies related to LLMs in Software Engineering (SE) \cite{hou:2024, shi:2025}, there is still a need for investigations focusing on managing the impact of LLMs. Mohamed et al. \cite{mohamed:2025} conducted a systematic literature review on how LLMs affect software developer productivity. From their thirty-seven selected studies, they summarised most benefits and risks concerning how LLMs affect software developer productivity, such as supporting knowledge acquisition and promoting over-reliance, disruptions to developer flow. They highlight that few studies explore aspects involving communication and human-human collaboration. Ferino et al. \cite{ferino:2026} conducted a systematic review exploring novice software developers' adoption and use of LLMs in SE activities. From their 80 selected studies, they identified many research gaps, such as exploring the impact of LLMs on mentorship interactions.

To gain more understanding about the impact of LLMs in SE tasks and especially how to manage it, our investigation focused on answering this main question: \textbf{\textit{How does the use of LLMs for software development impact software practitioners?}} This question was decomposed into the following research questions (RQs): 
{\renewcommand\labelitemi{}
\begin{itemize}[leftmargin=*]

    \item \textbf{RQ1}. \textit{How do LLMs help software developers move forward?} -- Moving developers forward involves the experienced benefits of using LLMs.

    \item \textbf{RQ2}. \textit{How do LLMs create an imbalance in software development?} -- Creating an imbalance involves experienced challenges when using LLMs. We draw special attention to this RQ in this report, as its findings may guide future studies.

    \item \textbf{RQ3}. \textit{How can software developers achieve a balanced use of LLMs?} -- It encompasses actionable recommendations to use LLMs.
\end{itemize}
}

To answer these RQs, especially RQ2, we conducted twenty-two semi-structured interviews with software practitioners across three rounds - between October (2024) and September (2025). Our study aims to comprehend the current industrial perspective on software developers adopting Large Language Model-based tools on SE-related activities, which involves exploring the benefits, challenges,  limitations, and recommendations shared by software practitioners involved in SE-related activities. Our analysis reveals the following main benefits: (B1) reduced effort due to LLMs as a foundation to boost code development and perceived saving time; (B2) flow experience when LLMs mitigate interruptions and automate simple and tedious tasks; (B3) interaction with  LLMs influencing developers' personality and creating a safe space; and (B4) LLMs promoting LLM entrepreneurship as a consultant for every (not complex) question; and these challenges: (D1) developer flow interruptions due to frictions in human-LLM interaction; (D2) reduced effort due to LLMs eroding developer skills and agency; (D3) interaction with LLMs reducing mentorship opportunities. The main contributions of this research include:

\begin{itemize}
    \item Identification and categorisation of benefits and challenges of adopting LLMs for software development tasks in terms of individual, team, organisation, and society level.
    \item A set of recommendations on how to best use LLMs for software development tasks.
\end{itemize}

\section{Methodology} \label{sec:methodology}

\begin{figure*}
    \centering
    \includegraphics[width=\textwidth]{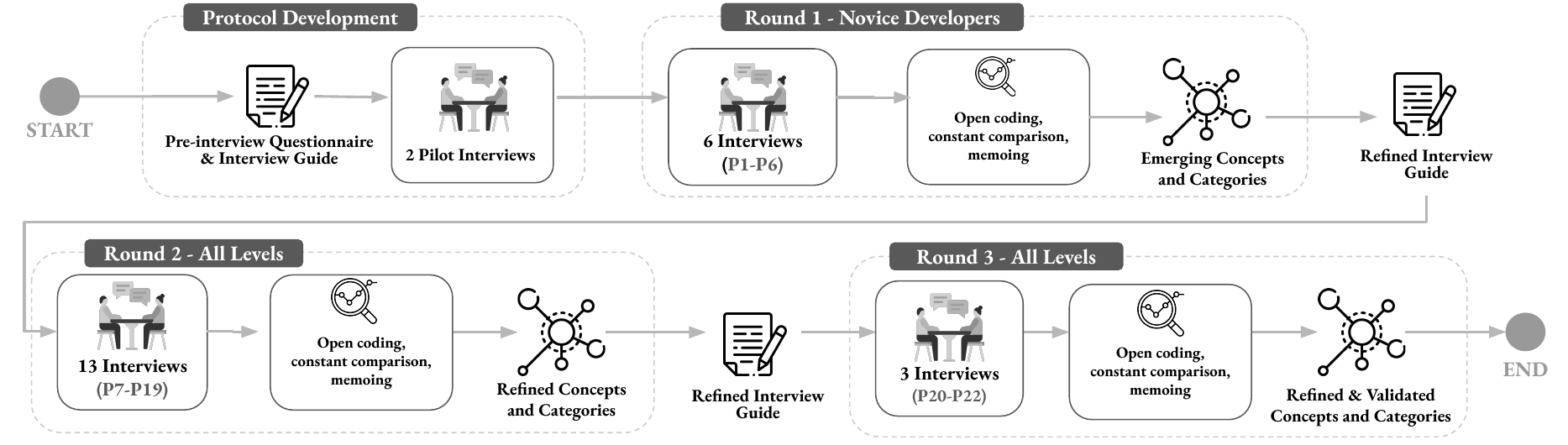}
    \caption{Iterative steps of data collection and analysis.}
    \label{fig:STGTStudyMethodology}
\end{figure*}


This study aims to comprehend the current industrial perspective on the impact of software practitioners adopting LLM-based tools for SE-related activities. We conducted an interview-based study with 22 software practitioners across three rounds of data collection\footnote{Approved by Monash Human Research Ethics Committee (No. 44875)}. We applied \textit{socio-technical grounded theory for data analysis} (STGT4DA) \cite{hoda:2022} to analyse the interview data iteratively. In our first round, we conducted six interviews with novice software practitioners. In our second round, we conducted interviews with thirteen novice (n=10) and experienced (n=3) software practitioners, resulting in the emergence of concepts and categories. Finally, we interviewed three more software practitioners during the third round of data collection and analysis, with the aim of consolidating our findings. Figure \ref{fig:STGTStudyMethodology} provides an overview of our study methodology. The supplementary material, which includes the pre-interview questionnaire, interview guide, STGT4DA coding example, assumption list, and LLM capabilities, is available online \cite{researchArtifact}.

\begin{table*}
\centering
\caption{Participant Demographics.}
\label{tab:participantDemographics}
\footnotesize
\begin{tblr}{
  row{even} = {c},
  row{3} = {c},
  row{5} = {c},
  row{7} = {c},
  row{9} = {c},
  row{11} = {c},
  row{13} = {c},
  row{15} = {c},
  row{17} = {c},
  row{19} = {c},
  row{21} = {c},
  row{23} = {c},
  cell{1}{1} = {c},
  cell{1}{2} = {c},
  cell{1}{3} = {c},
  cell{1}{4} = {c},
  cell{1}{5} = {c},
  cell{1}{6} = {c},
  hline{1-2,24} = {-}{},
}
\textbf{ID} & \textbf{Role} & \textbf{Domain} & \textbf{Country of Residence} & \textbf{Total Experience} & \textbf{Self-Reported Level of Experience} & \textbf{Round} \\
P1          & Software Developer       & IT              & Brazil           & 3-5 Years         & Experienced             & 1          \\
P2          & Software Engineer       & IT              & Australia           & 1-2 Years        & Experienced             & 1          \\
P3          & Software Engineer       & IT              & Australia           & 3-5 Years        & Intermediate          & 1          \\
P4          & Software Developer       & Finance         & Brazil           & 3-5 Years        & Intermediate          & 1          \\
P5          & Software Developer       & IT              & Australia           & 0-0.9 Years      & Intermediate          & 1          \\
P6          & Data Engineer       & DS/BD           & Brazil           & 3-5 Years        & Highly Experienced          & 1          \\
P7          & BI Analyst      & TELCOM          & Australia           & 0-0.9 Years      & Intermediate          & 2          \\
P8          & Software Engineer       & IT              & Australia           & 1-2 Years        & Experienced             & 2          \\
P9          & AI Engineer       & Government             & Brazil           & +10 Years        & Intermediate          & 2          \\
P10         & Software Engineer       & IT              & Canada           & 6-10 Years       & Highly Experienced          & 2          \\
P11         & Web Developer       & IT              & Australia           & 3-5 Years        & Highly Experienced          & 2          \\
P12         & Software Engineer       & Healthcare      & United States           & 6-10 Years       & Experienced             & 2          \\
P13         & Software Developer       & IT              & Australia           & 1-2 Years        & Advanced Beginner          & 2          \\
P14         & Data Engineer       & TELCOM          & Australia           & 1-2 Years        & Intermediate          & 2          \\
P15         & Researcher          & Engineering             & Australia           & 0-0.9 Years      & Novice           & 2          \\
P16         & R\&D Researcher       & Gaming          & Canada           & 1-2 Years        & Intermediate          & 2          \\
P17         & Data Analyst      & IT              & Australia           & 1-2 Years        & Intermediate          & 2          \\
P18         & Software Developer       & Finance         & Finland           & 3-5 Years        & Intermediate          & 2          \\
P19         & ML Scientist      & IT              & Canada           & 3-5 Years        & Experienced             & 2          \\
P20         & Software Engineer       & IT              & Malaysia           & 1-2 Years        & Intermediate          & 3          \\
P21         & Research Engineer     & IT              & Singapore           & 1-2 Years        & Experienced           & 3          \\
P22         & Software Developer       & IT              & United States           & +10 Years        & Highly Experienced          & 3          
\end{tblr}
\caption*{ \scriptsize{\textit{IT: Information Technology; DS/BD: Data Science/Big Data; TELCOM: Telecommunications.}}}
\end{table*}

\subsection{Study Design and Piloting} 

Based on the findings (i.e., benefits, challenges, recommendations) from our systematic literature review  \cite{ferino:2026}, we developed a preliminary interview guide, which also included potential follow-up questions. This interview guide \cite{researchArtifact} was improved based on discussions with the PhD supervisors, as well as feedback from an industrial collaborator, who has experience managing a software team in an Australian company. The \textit{Attitude Towards Artificial Intelligence} (ATAI) scale, developed by Sindermann et al. \cite{sindermann:2021},  was included in the pre-interview questionnaire to provide an initial understanding of participants' attitude towards AI, allowing the interviewer (first author) to adjust interview questions based on each participant's strongest responses across five dimensions — perceived benefits, trust, fear, job displacement, and existential concerns  — and also afforded additional contextual information to support the subsequent data analysis (i.e., open coding and memoing). ATAI scale includes the following five nine-point scale questions, which we adapted to a five-point scale ranging from \textsl{strongly disagree} to \textsl{strongly agree} to facilitate the participants to answer them: 

\begin{itemize}
    \item I fear artificial intelligence
    \item I trust artificial intelligence
    \item Artificial intelligence will destroy humankind
    \item Artificial intelligence will benefit humankind	
    \item Artificial intelligence will cause many job losses
\end{itemize}

From our industry collaborator's feedback, we included questions, for example, seeking to understand participants' perceptions of the impact of LLMs on their career trajectory or job market competitiveness in the near future. We collected participants' information regarding basic demographics, work experience, and experience with LLM tools. We also conducted two pilot interview sessions with experienced software practitioners to evaluate the interview structure - whether to adopt a pre-interview questionnaire to collect participants' information - and the clarity of the interview questions. During our pilot study, we observed that it takes only about 5-10 minutes to collect participants' information during the interview; however, we also observed that the pilot study participant who filled out a pre-interview questionnaire was more comfortable (relaxed) during the interview. We believe that while using the pre-interview questionnaire, with the interview moment focused only on the interview guide, we provided a more simplified structure for the participant. The pilot interview using a pre-interview questionnaire was not included in the analysis because this participant had no experience using LLMs for software development.

\subsection{Sampling}

 Leveraging the findings from our SLR \cite{ferino:2026} on novice developers' adoption of LLMs for SE tasks as a foundation, we began with purposive sampling in the first round, focusing on less experienced developers with less than five years of professional experience (See Fig. \ref{fig:STGTStudyMethodology}). In the following rounds, we moved to convenience sampling, interviewing software practitioners from all levels of experience since our first round of data collection and analysis did not result in enough codes, concepts, or memos to justify continuing only with novice participants. This would also support further comparison between novice and experienced developers' perceptions and experiences. We advertised this study on our professional social media, LinkedIn and X (formerly Twitter). We also recruited participants from our personal connections. Initially, we advertised our study, including a \$50 (AUD) voucher gift card. However, due to potential fake participants identified based on discrepancies between their IP addresses and the countries they submitted in the pre-interview questionnaire, we got approval from our Faculty Ethical Review Committee to omit the gift card in the advertisement, and only offered it to genuine participants at the end of the interview. We also adopted a snowballing approach by encouraging the participants to share our study with others.

\begin{table*}
    \caption{Participants' use of LLMs.}
    \label{tab:extraDemographicInfo}
    \resizebox{\textwidth}{!}{
    \begin{tabular}{@{}llllll@{}}
        \toprule
        \textbf{Exp. (LLMs in General)} & \textbf{\# of Practitioners} & \textbf{Exp. (LLM4SE)}& \textbf{\# of Practitioners} & \textbf{\# of Prompts per Day} & \textbf{\# of Practitioners}\\
        \midrule
        
         Less than 1 year & \blackbar{5}{5}     & Less than 6 months & \blackbar{4}{4}      & Less than or equal to 2 & \blackbar{2}{2}\\
         Less than 1.5 year  & \blackbar{5}{5} & Less than 1 year  & \blackbar{8}{8}       &  3 - 5  & \blackbar{5}{5}\\
        Less than 2 years & \blackbar{7}{7}     & Less than 1.5 year & \blackbar{3}{3}      & 6 - 10 & \blackbar{3}{3}\\
        More than 2 years & \blackbar{5}{5}     & Less than 2 years & \blackbar{5}{5}     & 11 - 20; 21 - 30 & \blackbar{4}{4} \\
                        &                       & More than 2 years & \blackbar{2}{2}    & 50 - 100; +100 & \blackbar{2}{2}\\
         
        \cmidrule(r){1-4} \cmidrule(r){5-6}
        \multicolumn{2}{@{}l}{\textbf{ATAI Scale}} & \multicolumn{2}{@{}l}{\textbf{Frequency*}} & \textbf{SE Tasks supported by LLMs} & \textbf{\# of Practitioners}\\
        \cmidrule(r){1-4} \cmidrule(r){5-6}

        \multicolumn{2}{@{}l}{I fear artificial intelligence} & \multicolumn{2}{@{}l}{\multicolorbar{7}{7}{4}{2}{2}{[7, 7, 4, 2, 2]}} & Code-related tasks & \blackbar{20}{20}\\
        
        \multicolumn{2}{@{}l}{I trust artificial intelligence} & \multicolumn{2}{@{}l}{\multicolorbar{4}{10}{4}{4}{0}{[4, 10, 4, 4, 0]}}  & Test-related tasks & \blackbar{11}{11}\\
        
        \multicolumn{2}{@{}l}{Artificial intelligence will destroy humankind} & \multicolumn{2}{@{}l}{\multicolorbar{11}{3}{3}{4}{1}{[11, 3, 3, 4, 1]}} & Documentation-related tasks & \blackbar{9}{9}\\
        
        \multicolumn{2}{@{}l}{Artificial intelligence will benefit humankind} & \multicolumn{2}{@{}l}{\multicolorbar{1}{0}{3}{12}{6}{[1, 0, 3, 12, 6]}} & Requirement-related tasks  & \blackbar{7}{7}\\
        \multicolumn{2}{@{}l}{Artificial intelligence will cause many job losses} & \multicolumn{2}{@{}l}{\multicolorbar{0}{2}{7}{10}{3}{[0, 2, 7, 10, 3]}} &  & \\

        \cmidrule(r){1-4} \cmidrule(r){5-6}
        \textbf{LLM tools} & \textbf{\# of Practitioners} & \textbf{LLM tools} &  \textbf{\# of Practitioners} & \textbf{Company's AI Policy} & \textbf{\# of Practitioners}\\
        \cmidrule(r){1-4}\cmidrule(r){5-6}

        ChatGPT & \blackbar{18}{18} &  Cursor & \blackbar{2}{2} & Allowed to use  & \blackbar{16}{16}\\
        
        GitHub Copilot &\blackbar{7}{7} &  Phind & \blackbar{2}{2} & No policy & \blackbar{3}{3}\\
        
        Claude & \blackbar{7}{7} &  Mistral & \blackbar{2}{2} &  Prohibited to use & \blackbar{1}{1}\\
        
        Llama & \blackbar{4}{4} & Jan.AI & \blackbar{1}{1} &   I do not know & \blackbar{1}{1}\\

         Gemini & \blackbar{2}{2} & H2O Danube & \blackbar{1}{1}  &   Prefer not to say & \blackbar{1}{1} \\
            
         Perplexity & \blackbar{2}{2}  &  Continue & \blackbar{1}{1}  &    &  \\
            
         Microsoft Copilot & \blackbar{2}{2}  &  &  &   &  \\
            
        \bottomrule
    \end{tabular}
    }

    \begin{flushleft}
         \textit{*Order of the bars in Frequency column of ATAI Scale graph: Strongly disagree → Somewhat disagree → Neither agree nor disagree → Somewhat agree → Strongly agree, followed by respective values. }
    \end{flushleft}

\end{table*}

The first round of interviews with six novice software practitioners took place between October and November (2024). Then, after changes in the interview guide, the second round of interviews with thirteen software practitioners took place between April and June (2025). The third round of interviews occurred in September (2025). The interviews were conducted and recorded via Zoom meetings, which also provided the transcriptions. Interviews were scheduled to go 40-45 minutes, but ranged from 34 to 58 minutes. We manually reviewed the transcriptions, filtered any sensitive information, and de-identified the participants.

\textbf{Participants' demographics.} Our participant pool covers a diverse range of roles, countries of residence, years of experience, and experience using LLMs. We conducted twenty-two interviews with software practitioners from Latin America, Oceania, North America, Asia, and Europe - mostly from Australia (n=10). Table \ref{tab:participantDemographics} provides an overview of participants' demographics. Our study participants worked in different domains, such as healthcare, government, and finance. Most of the participants are males (n=14), have equal to or less than five years of professional experience (n=13), and describe their skill level as intermediate or advanced beginner (n=17).

Regarding their experience using LLM tools, which can influence the faced challenges and suggested solution strategies, participants reported using a variety of LLM tools. Not surprisingly, ChatGPT (n=18) was the most recurrent tool used by the interview participants. Table \ref{tab:extraDemographicInfo} summarises participants' demographics information related to the adoption of LLM tools. Most of the participants (n=14) have used LLMs in general, not specifically for SE tasks,  for more than one year. According to their self-report amount of daily prompts, 59\% (n=13) of participants engage daily with LLMs through at least 6 prompts. Their attitudes towards LLMs captured via the \textit{Attitude Towards Artificial Intelligence} (ATAI) scale highlight different inclinations towards AI technologies. While mostly (n=14) \textsl{strongly disagreeing} or \textsl{somewhat disagreeing} about fearing AI, they also mostly (n=14) \textsl{disagree} or \textsl{somewhat disagree} about trusting AI. Most (n=18) \textsl{somewhat agree} or \textsl{strongly agree} that AI will benefit humanity, and \textsl{somewhat disagree} or \textsl{strongly disagree} that AI will destroy humanity (n=14). However, most participants (n=13) \textsl{somewhat agree} or \textsl{strongly agree} that AI will affect the job market, leading to job losses. With respect to their experience using LLMs for SE tasks, mostly (n=16) reported having more than six months of experience, and they use LLMs mostly for code-related tasks (n=17), such as coding and debugging. Most of the study participants (n=14) reported that their companies allow them to employ LLM tools for job responsibilities.

The fast-paced evolution of LLM tools compels us to examine the features available in the tools mentioned by interview participants. During our three rounds of data collection and analysis, we observed an emergence and evolution of different LLM tools (e.g., ChatGPT, Cursor) and integration with traditional tools (e.g., ChatGPT for Databricks \cite{databricksgpt5:2025}, Copilot for Power BI \cite{mscopilot:2025}). This was also highlighted by the participants, e.g.: \faComments \;\textit{``I've been able to see a bit of how those tools evolve during this time. [For example, GitHub] Copilot evolved from being a better autocomplete to having more tools [and features]."} -- P10 [Software Engineer]. Ferino et al. \cite{ferino:2026} suggest that researchers investigating LLMs for SE should provide a snapshot of the LLM features with the intent of improving clarity of the context related to the data collection. We summarise the main features in the supplementary online package \cite{researchArtifact}.

\begin{figure*}
    \centering
    \includegraphics[width=\textwidth]{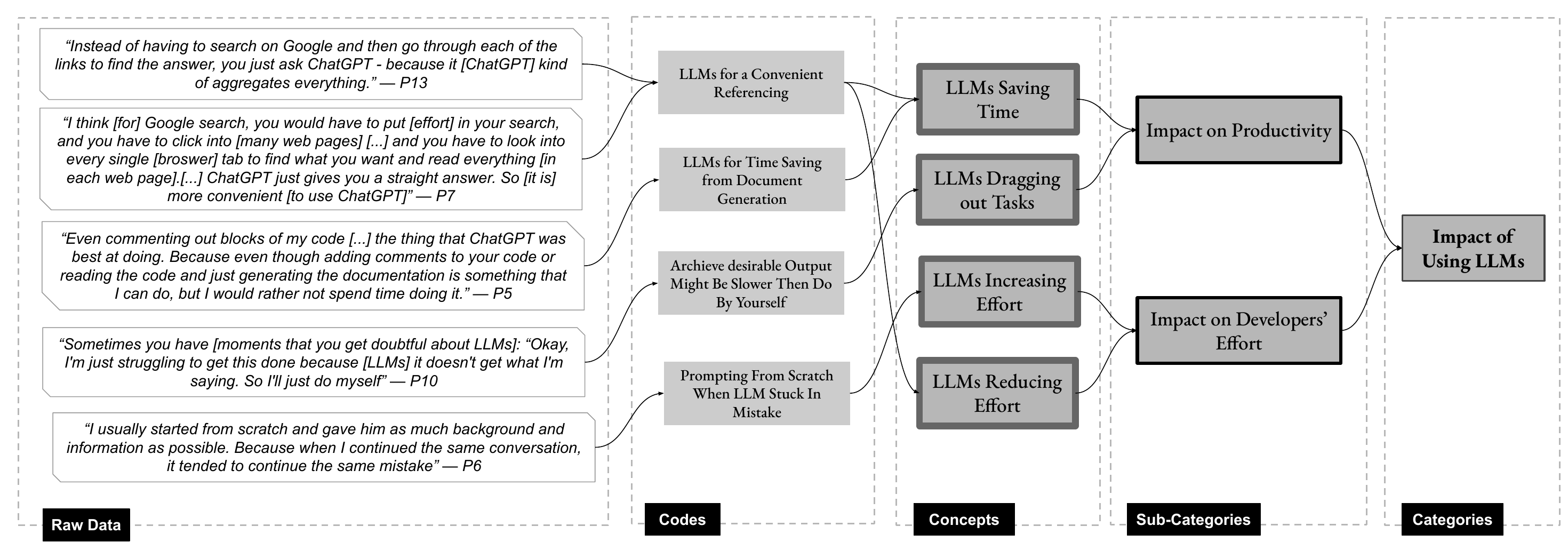}
    \caption{Emergence of the category ``\textsl{Impact on using LLMs}" from \textit{raw data} → \textit{codes} → \textit{concepts} →
\textit{subcategories} → \textit{category} through constant comparison \cite{hoda:2024}.}
    \label{fig:OpenCodeConstantComparison}
\end{figure*}

\subsection{Data Analysis}

Applying STGT's \textit{basic stage} of data collection and analysis, we conducted a lean literature review to identify research gaps, followed by data collection (e.g., survey), and hashtag open coding, constant comparison, and basic memoing for data analysis \cite{hoda:2022, hoda:2024}. STGT4DA has been applied in various SE studies to identify \textit{``important patterns in the qualitative data and present them as layered and/or multi-dimensional findings, along with insights and reflections"} \cite{pant:2024, madampe:2024, madugalla:2024, gunatilake:2025}. We leveraged its foundational socio-technical research framework to investigate the \textit{socio-technical phenomenon} of software practitioners' perspectives on the role of LLM tools in software development, in the \textit{socio-technical domain} of LLMs for SE (LLM4SE), with \textit{socio-technical actors} such as software developers, software engineers, data engineers, data scientists, and DevOps engineers. As \textit{socio-technical researchers}, we combined our skills and experiences. The interviews were conducted by an early-career researcher (first author) under the guidance of three experienced researchers, and the data were analysed by the first author with detailed feedback from the second author. We also applied \textit{socio-technical data, tools, and techniques} such as a Qualtrics survey to collect participant demographics, Zoom for recording and transcription, DeepL\footnote{\url{www.deepl.com}} for translation of the interview transcriptions in Portuguese (P1, P4, P6) to English, and Nvivo and spreadsheets to support the analysis of the interview transcriptions.

\textbf{Hashtag Open Coding.} The process of hashtag open coding \cite[Chapter 10]{hoda:2024} was iteratively improved by the first researcher following the guidance of the second author. Initially, the first author printed out one of the interviews and coded it, followed by a review and discussions with an experienced researcher. After this, the researcher coded two more interviews using Google Docs. The researcher decided to move to NVivo, instigated by its features, and spreadsheets. Before performing the hashtag open coding, and between rounds of data collection and analysis, the first author was advised to create and update an assumption list \cite{researchArtifact}. This helped to make the researcher's assumptions and biases transparent so they could be managed. 

\textbf{Constant Comparison.} Similar codes were grouped into concepts, and similar concepts under sub-categories, and similar sub-categories under categories. During this process, we drew diagrams to facilitate visualisation and insights regarding the relationship between concepts and sub-categories, and also conducted discussions between the authors. Figure \ref{fig:OpenCodeConstantComparison} illustrates the emergence of the category \textit{Impact of using LLMs} via coding and constant comparison.

\textbf{Memoing.} While performing constant comparison, we also wrote down memos reflecting our ideas, thoughts, and key elements emerging throughout the analysis. Those memos, which encompass both interview and inter-interview levels, supported the update of the interview guide between rounds. At the interview level, we compared the participants' pre-interview questionnaire responses with their interview responses, especially concerning the ATAI scale. That information provided more context for our analysis. At the inter-interview level, we compared codes from different interview participants, which also helped to refine our concepts and sub-categories. An example of a memo ``How do LLMs affect developer intuition?", is shown below. The discussion on the main insights from memoing is presented in Section \ref{sec:implicationsToResearch}.

\begin{tcolorbox}[float=htpb!,colback=gray!5!white,colframe=gray!75!black,title={\textbf{Memo on \textit{``How do LLMs affect developer intuition?"}}}]
LLMs seem to emerge as a learning tool for software developers based on participants P12, P14, and P15, e.g., relying on LLMs to cover gaps in knowledge. At the same time, P14 argues that while using LLMs for learning, novice developers should do the majority of the work. This practice seems to be necessary to improve novice developers' skills, but also to cultivate the intuition inherent in experienced developers. Intuition is something that comes naturally after fully understanding something.  That intuition (internalised knowledge) helps to avoid scenarios where LLMs are misleading, which P11 described as ChatGPT being \textit{``a very good liar''}.    
\end{tcolorbox}

\section{Findings} \label{sec:findings}

We found variations in the impact of software developers adopting LLMs across four levels: 

\begin{itemize}
    \item \textit{\textbf{Impact on the Individual Level}}: it refers to LLMs impacting software practitioners directly;
    \item \textit{\textbf{Impact on the Team Level}}: it involves how LLMs influence software development teams and their collaboration;
    \item \textit{\textbf{Impact on the Organisation Level}}: it encompasses how LLMs affect entire software organisations;
    \item \textit{\textbf{Impact on the Society Level}}: it refers to how LLMs affect daily life in society, not just developer communities.
\end{itemize}

\begin{figure*}
    \centering
    \includegraphics[width=0.75\textwidth]{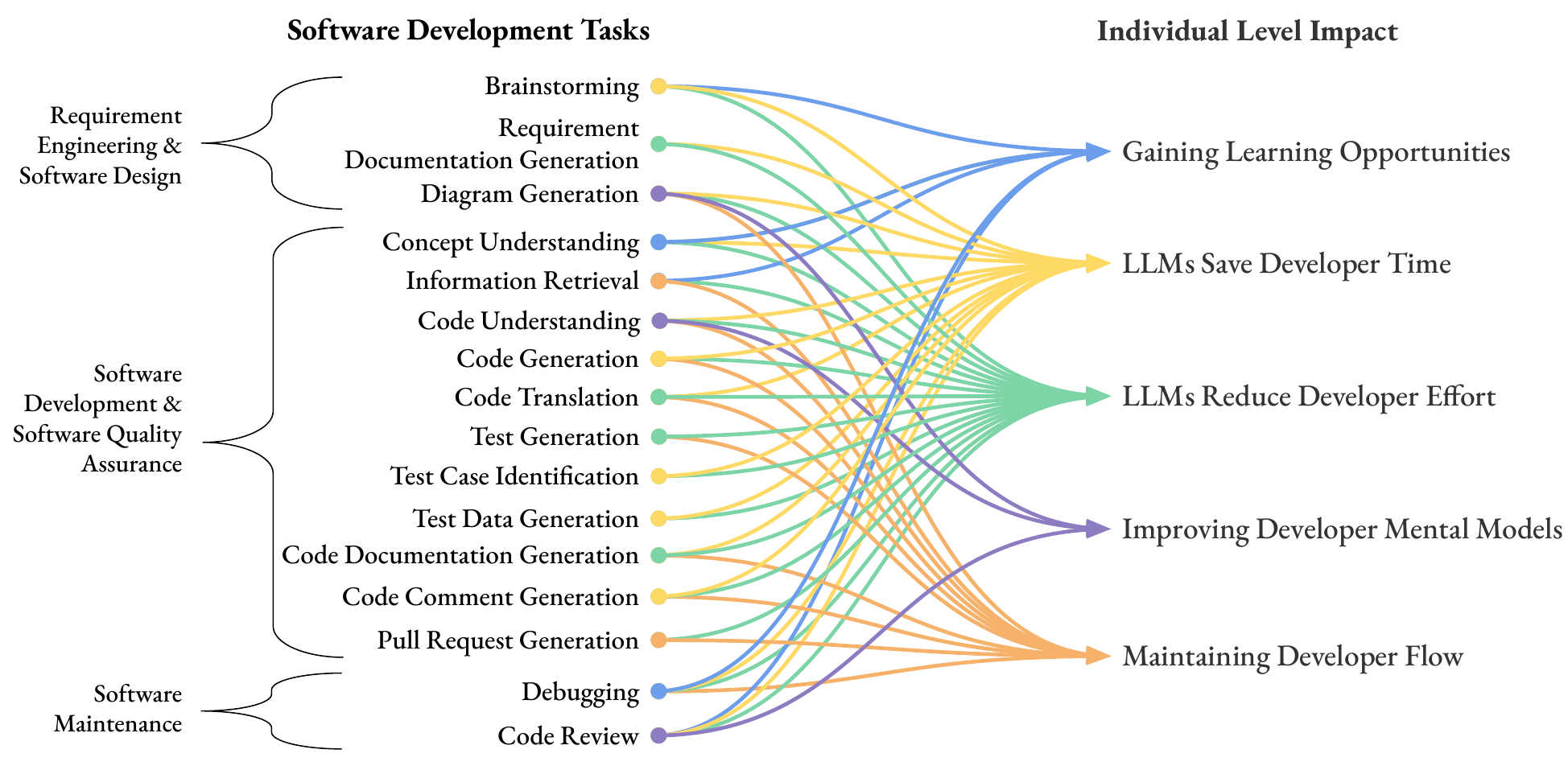}
    \caption{Main Benefits of using LLMs at the Individual (Software Practitioner) Level.}
    \label{fig:MainAdvantagesIndividualLevel}
\end{figure*}

\begin{figure*}
    \centering
    \includegraphics[width=0.75\textwidth]{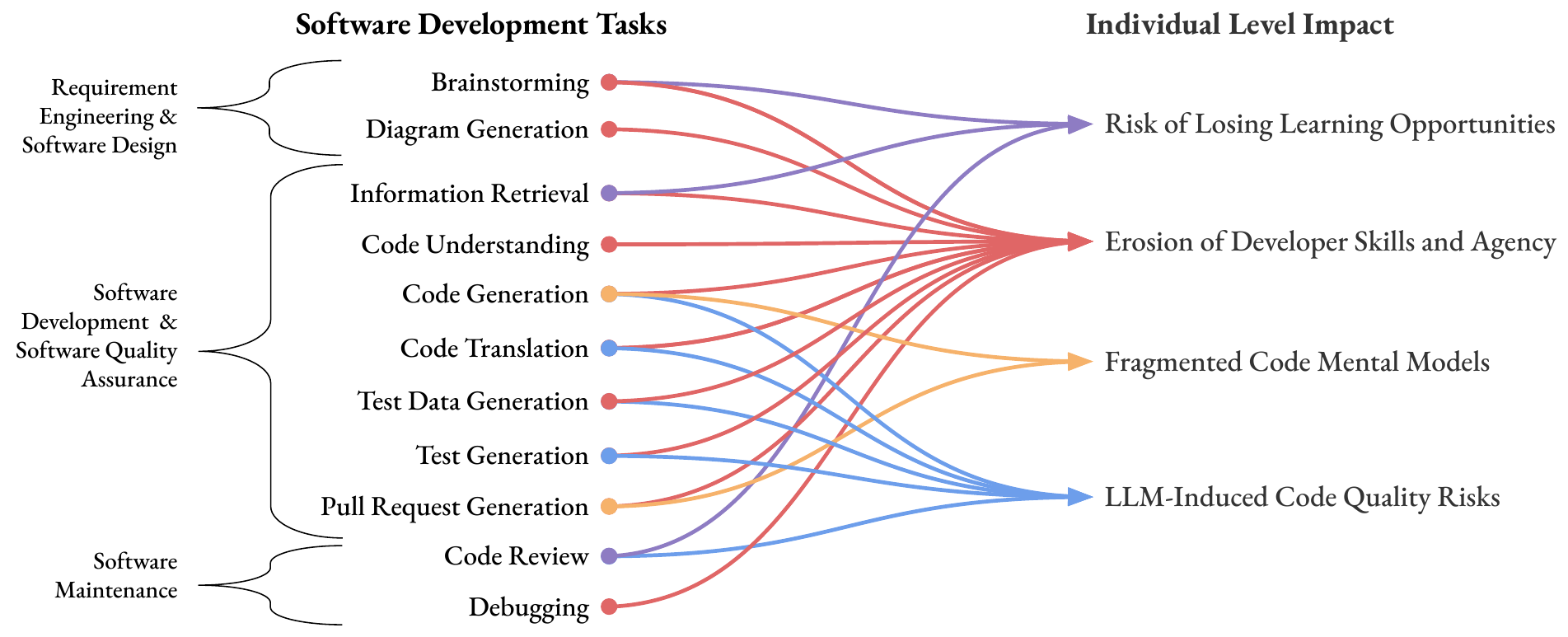}
    \caption{Main Challenges of using LLMs at the Individual (Software Practitioner) Level.}
    \label{fig:MainDisadvantagesIndividualLevel}
\end{figure*}

Tables \ref{tab:benefitsIndividualLevel} and \ref{tab:disadvantageIndividualLevel} illustrate the benefits and challenges at the individual level of using LLM tools to support software development tasks, along with the \textsl{reasons} influencing these impacts. For instance, Table \ref{tab:disadvantageIndividualLevel} shows that using LLMs for brainstorming may hinder developers’ skills due to a lack of practice in software development skills and overreliance on LLMs. Additionally, we have summarised the main benefits and challenges in Figures \ref{fig:MainAdvantagesIndividualLevel} and \ref{fig:MainDisadvantagesIndividualLevel}, respectively. For instance, using LLMs for debugging can boost code development, saving development time, reducing software developers' effort, and providing learning opportunities to software developers.  We organised the remainder of this section by describing the\textit{benefits of using LLMs} (section \ref{subsec:benefits}), \textit{challenges when using LLMs} (section \ref{subsec:disadvantages}), and \textit{recommendations on how to use LLMs} (section \ref{subsec:recommendations}). Throughout this section, we provide representative quotes to illustrate participants' experiences and viewpoints. In order to protect the privacy of our participants, we refer to the interview participants as P1-P22. Similar to Masood et al.\cite{masood:2020}, we will use some expressions to indicate the extent of the predominance. In this sense, `few' refers to less than 25\%, `many or majority' to over 50\%. Although this is not a quantitative study, it indicates how much evidence supports each theme.

\subsection{RQ1. How do LLMs help software developers move forward?} \label{subsec:benefits}

\subsubsection{Individual Level}

\textbf{Reducing developer effort with LLMs.} A majority of participants (i.e., P1-P18, P20-P22) described scenarios where LLMs work to lighten their efforts for simple, repetitive, or tedious tasks. This happens because software practitioners hand over the control to LLMs' automation capabilities, e.g.: \faComments \;\textit{``As I have this extension in Databricks, it makes it much easier. For example, [if] I miss a line, I miss a path, I miss a way of calling a class method, etc, it corrects it on the spot and automatically shows the button: `Do you want me to correct it and run the cell again?'."} -- P6 [Data Engineer]. \label{subsubsec:repetitiveTasks}

\textbf{Saving developer time with LLMs.} Most of the participants (i.e., P1-P18, P20, P22) also describe how LLMs contribute to time-saving for simple, repetitive, or tedious tasks. The perceived saving in developers' time happens due to the reduced effort provided by LLMs when the developer hands over control to LLMs, e.g.: \faComments \;\textit{``In the past, there were several sites, Stack Overflow [...] From the moment you go looking for your question until you find it, you've already spent a lot of time."} -- P6 [Data Engineer].

\textbf{Gaining learning opportunities.} Fourteen participants (i.e., P1-P4, P6-P7, P12, P14, P15, P17, P18, P20, P22) mentioned the potential of LLMs as an educational tool for software developers, as well as the learning opportunities from adopting LLMs. For instance, developers can search for the reason behind AI's wrong suggestion, e.g.: \faComments \;\textit{``I try to see if the wrong answer was because I didn't give enough information [...] I'll see if the prompt didn't make sense"} -- P1 [Software Developer]. LLMs can provide alternative solutions that can reinforce developers' solution arsenal, e.g.: \faComments \;\textit{``LLMs will give me an essentially different perspective. Maybe this is an alternative approach, a more concise way of doing things"} -- P17 [Data Analyst]. \label{subsubsec:exploratoryLearning}

\begin{table*}
\centering
\caption{Summary of the Benefits of using LLMs at the Individual (Software Practitioner) Level.}
\label{tab:benefitsIndividualLevel}
\footnotesize
\begin{tblr}{
width = \linewidth,
  colspec = {Q[86]Q[304]Q[344]},
  column{1} = {r},
  cell{1}{1} = {c},
  cell{2}{1} = {r=3}{},
  cell{5}{1} = {r=6}{},
  cell{11}{1} = {r=2}{},
  vlines,
  hline{1,13} = {-}{0.08em},
  hline{2,5,11} = {-}{},
  hline{3-4,6-10,12} = {2-3}{},
}
\textbf{\textbf{SE Activity}}                        & \textbf{\textbf{Software Development Tasks / How LLMs Impact?}}                                                                                                                                                                                                                                               & \textbf{Why the Impact Happens?}                                                                                                                                                                   \\
{Requirement Engineering \&\\  Software Design}        & {\textbf{\textit{Task}}\textit{:}~Brainstorming\\\textbf{\textit{Benefit}}\textit{:}~[\#1] Reducing effort, [\#2] LLMs saving time, [\#3] Gaining learning opportunities}                                                                                                    & \textit{\textbf{Reason:~}} ~LLMs Reducing $\text{Effort}^{\#2}$;~~Mitigating  $\text{interruptions}^{\#1,\#2}$; Handing over the Control to $\text{LLMs}^{\#1,\#2}$; $\text{Automation}^{\#1,\#2}$; Reliance on $\text{LLMs}^{\#1,\#2, \#3}$; Balanced control over $\text{LLMs}^{\#3}$; LLMs $\text{capabilities}^{\#3}$      \\

& {\textit{\textbf{Task}:} Requirement Documentation Generation\\\textbf{\textit{Benefit: }}[\#1] Reducing effort,{[}\#2] LLMs saving time~}                                                                                                                                 & \textbf{\textit{Reason:}}~LLMs reducing effort$\text{}^{\#2}$; Mitigating interruptions$\text{}^{\#1,\#2}$; Handing over the control to LLM$\text{}^{\#1,\#2}$; Automation$\text{}^{\#1,\#2}$; Reliance on LLMs$\text{}^{\#1,\#2}$                                            \\
                                                     
& {\textit{\textbf{\textbf{Tasks}}:}~Diagram Generation\\\textit{\textbf{Benefit: }}[\#1] Reducing effort,{[}\#2] LLMs saving time, [\#3] Improving developer mental models, [\#4] Maintaining developer flow}                                                              & \textbf{\textit{Reason:~}}LLMs reducing effort$\text{}^{\#2,\#4}$; Mitigating interruptions$\text{}^{\#1,\#2, \#4}$; Handing over the control to LLM$\text{}^{\#1,\#2, \#3, \#4}$; Automation$\text{}^{\#1,\#2, \#3, \#4}$; Reliance on LLMs$\text{}^{\#1,\#2, \#3, \#4}$                                          \\

{Software Development \& \\Software Quality Assurance} & {\textit{\textbf{\textbf{Tasks}}:}~Concept Understanding\\\textit{\textbf{Benefit: }}[\#1] Reducing effort, [\#2] LLMs saving time, {[}\#3] Gaining learning opportunities}                                                                                                                                      & \textbf{\textit{Reason:}}~Balanced control$\text{}^{\#3}$; Reliance on LLMs$\text{}^{\#1,\#2,\#3}$; LLMs capabilities$\text{}^{\#3}$; Mitigating interruptions$\text{}^{\#1,\#2}$; Automation$\text{}^{\#1,\#2}$; LLMs reducing effort$\text{}^{\#2}$                                                                                                                       \\

& {\textit{\textbf{\textbf{Tasks}}:}~Information Retrieval\\\textbf{\textbf{\textit{Benefit: }}}[\#1] Reducing effort, [\#2] LLMs saving time, [\#3] Gaining learning opportunities, [\#4] Maintaining developers' flow}                                                         & \textbf{\textit{Reason:~}}LLMs saving time$\text{}^{\#1}$; LLMs reducing effort$\text{}^{\#2,\#4}$; Mitigating interruptions$\text{}^{\#1,\#2,\#4}$;~Handing over the control to LLM$\text{}^{\#1,\#2,\#4}$; Automation$\text{}^{\#1,\#2,\#4}$; Reliance on LLMs$\text{}^{\#1,\#2,\#3,\#4}$; Balanced control$\text{}^{\#3}$; LLMs capabilities$\text{}^{\#3}$       \\

& {\textit{\textbf{\textbf{Tasks}}:}~Code Understanding\\\textbf{\textbf{\textit{Benefit: }}}[\#1] reducing effort, {[}\#2] LLMs save developer time, [\#3] Improving developer mental models, {[}\#4] Maintaining developer flow}                                                  & \textbf{\textit{Reason:} }LLMs reducing effort$\text{}^{\#2,\#4}$; Mitigating interruptions$\text{}^{\#1,\#2,\#4}$; Handing over the control to LLM$\text{}^{\#1,\#2,\#3,\#4}$; Automation$\text{}^{\#1,\#2,\#3,\#4}$; Reliance on LLMs$\text{}^{\#1,\#2,\#3,\#4}$                                             \\

& {\textit{\textbf{\textbf{Tasks}}:}~Code Generation; Code Translation; Code Documentation Generation; Code Comment Generation;~ Unit Test Generation\\\textbf{\textbf{\textit{Benefit: }}}[\#1] Reducing effort, {[}\#2] LLMs saving time, [\#3] Maintaining developers' flow} & {\textit{\textbf{Reason:}}~Reducing effort$\text{}^{\#2,\#3}$; Mitigating interruptions$\text{}^{\#1,\#2,\#3}$; Handing over the control to LLM$\text{}^{\#1,\#2,\#3}$; Automation$\text{}^{\#1,\#2,\#3}$; Reliance on LLMs$\text{}^{\#1,\#2,\#3}$}                                          \\

& {\textit{\textbf{\textbf{Tasks}}:}~Test Case Identification,~Test Data Generation\\\textbf{\textbf{\textit{Benefit:~}}}[\#1] Reducing effort, {[}\#2] LLMs saving time~\textit{\textbf{\textbf{~}}}}                                                                          & {\textit{\textbf{Reason:}}~ Reducing effort$\text{}^{\#1}$; Mitigating interruptions$\text{}^{\#1,\#2}$; Handing over the control to LLM$\text{}^{\#1,\#2}$; Automation$\text{}^{\#1,\#2}$; Reliance on LLMs$\text{}^{\#1,\#2}$}                                          \\
                                                     
& {\textit{\textbf{\textbf{Tasks}}:}~Pull Request Generation\\\textbf{\textbf{\textit{Benefit: }}}[\#1] Reducing effort, {[}\#2] LLMs saving time}                                                                                          & \textbf{\textit{Reason:~}}Mitigating interruptions$\text{}^{\#1,\#2}$;~Cautious reliance on LLMs$\text{}^{\#1,\#2}$; Reducing effort$\text{}^{\#2}$; LLMs saving time$\text{}^{\#1,\#2}$; Balanced control$\text{}^{\#1,\#2}$; Handing over the control to LLM$\text{}^{\#1,\#2}$; Automation$\text{}^{\#1,\#2}$; Reliance on LLMs$\text{}^{\#1,\#2}$;   \\

Software Maintenance                                 & {\textit{\textbf{\textbf{Tasks}}:}~Debugging\\\textbf{\textbf{\textit{Benefit: }}}[\#1] Reducing effort, {[}\#2] LLMs saving time,[\#3] Gaining learning opportunities,\\{[}\#4] Maintaining developers' flow}                                                                & {\textit{\textbf{Reason:} } LLMs reducing efforts$\text{}^{\#2,\#4}$; Mitigating interruptions$\text{}^{\#1,\#2, \#4}$; Handing over the control to LLM$\text{}^{\#1,\#2,\#4}$; Balanced control$\text{}^{\#3}$; LLMs capabilities$\text{}^{\#3}$; Automation$\text{}^{\#1,\#2,\#4}$; Reliance on LLMs$\text{}^{\#1,\#2,\#3, \#4}$} \\

& {\textit{\textbf{\textbf{Tasks}}:}~Code Review\\\textbf{\textbf{\textit{Benefit: }}}[\#1] Reducing effort, {[}\#2] LLMs saving time, [\#3] Gaining learning opportunities, {[}\#4] Improving developer skills}                                                             & {\textbf{\textit{Reason:}} LLMs saving time$\text{}^{\#1,\#2}$; LLMs reducing effort$\text{}^{\#1,\#2}$; Balanced control$\text{}^{\#1,\#2}$; Automation$\text{}^{\#1,\#2}$; Cautious reliance on LLMs$\text{}^{\#1,\#2}$}                                                                         
\end{tblr}
\end{table*}

\textbf{Maintaining developer flow.} Fifteen participants (i.e., P1, P4-P8, P12-P13, P15-P20, P22) presented different aspects that LLMs contribute to the software development flow. For instance, LLMs help developers to stay focused on the logic instead of typing when it autocompletes, e.g.: \faComments \;\textit{``So it helps me with  [...] reducing the time I'm actually writing and giving me more space to think about the flow in general."} -- P4 [Software Developer].

\textbf{Improving developer mental models.} A few participants (i.e., P6, P10, P13-P14)  mention how LLMs can aid developers in improving their code understanding, e.g.: \faComments \;\textit{```Why is it going on? Why is that method function [implemented in that way]?'"} -- P19 [Machine Learning Scientist].

\textbf{Safe space.} A few participants (i.e., P4, P7, P12, P15) mention how LLMs provide a safe space for sharing ideas and questions. For instance, software developers can ask as many questions as they need, e.g.: \faComments \;\textit{``I feel like senior people don't have a lot of time on hand for trivial issues [...]  [but for LLMs] there is no judgment [while asking questions], because you can ask dumb questions to ChatGPT or any [other] LLMs. And you can ask it as many times as you want, and you can get examples [...] You can get customised responses."} -- P15 [Researcher]. This seems to be motivated by the desire to avoid disturbing colleagues' work, while becoming self-reliant, e.g.: \faComments \;\textit{``Everyone's very busy at work - the senior developers as well - you kind of - It's not a set rule - but I feel I have a limited number of questions I can ask per day."}  -- P7 [BI Analyst]. In parallel, LLMs can offer certain levels of companionship experience, e.g.:  \faComments \;\textit{``We can interact with LLMs like a person. And it's good to have that kind of companionship as well, like you can talk to it and say: `Hey, what do you think about this?'"} -- P12 [Software Engineer], and empathy, e.g.: \faComments \;\textit{``There was one time when [...] I imported some of the things Copilot sent me and tried them. When I went to Unity, it wouldn't even run the project. [...]  [Copilot would reply]: `I'm sorry for your dissatisfaction, for your frustration, but let's try to do it another way'."} -- P4 [Software Developer].

\subsubsection{Team Level} \label{subsec:teamLevel}

\textbf{Mitigating interruptions.} Fifteen participants (i.e., P1, P4-P6, P10, P12-P21) described how LLMs lead to fewer interruptions. Participants showed a preference to LLMs over disturbing colleagues, e.g.: \faComments \;\textit{``I'm always more comfortable using ChatGPT, Copilot. I do as little as possible to ask for help, so as not to disturb other people."} -- P4 [Software Developer]. Participants also highlight the potential of LLMs supporting companies' onboarding, which normally requires a lot of guidance from colleagues,  e.g.: \faComments \;\textit{``If you're kind of new to the team, you would feel more inclined to kind of just break it down through Copilot or ChatGPT first. And then afterwards, if you still don't understand it, then kind of go ask an experienced coworker"} -- P14 [Data Engineer]. \label{subsubsec:onboarding}

\subsubsection{Organisation Level} \label{subsec:OrganisationLevel}
\textbf{Cost savings due to LLMs.} A few participants (i.e., P7, P9, P15-P16, P20-P22) highlight how LLMs can contribute to reducing costs. For instance, using LLMs for debugging can decrease the time to find the problem, e.g.: \faComments \;\textit{``We are a small company, but we move at least one to 2 million dollars per week [...] If there's a production issue, then it's like: `Oh, we need to solve this immediately'. And it [LLMs] saves a lot of time to find the bugs and to mitigate the problem. That [time spent searching for the problem] it costs, [and] you have put the money in somewhere else, where it should not be"} -- P12 [Software Engineer].

\subsubsection{Societal Level} \label{subsec:SocietyLevel}

\textbf{Foster Entrepreneurship.} A few participants (i.e., P10 and P21) mention the potential of LLMs to encourage entrepreneurship. This occurs due to the great LLM prototyping capabilities, e.g.: \faComments \;\textit{``Prototyping becomes much faster for these [startup] companies [...] Even a non-coder [...] can create just the rundown prototype, not a full, scalable solution. But just to showcase sort of a working front-end"} -- P21 [Researcher Developer], and boost in individual productivity, reducing demand for big teams, e.g.: \faComments \;\textit{``It also makes it easier to create new companies [...]  because you can do more with fewer people.} -- P10 [Software Engineer]. At the same time, LLMs can provide basic information necessary to start a business, e.g.: \faComments \;\textit{``It can help you with the basics of marketing, accounting, and law, so it can help you tap into the basics of all these things, so you don't need to recruit a team first."} -- P21 [Researcher Developer].

\begin{tcolorbox}[float=htpb!,colback=black!5!white,colframe=black!75!black,title={\textbf{RQ1.} \textit{How do LLMs help software developers move forward?}}]
Our findings reveal benefits at the individual ($n$=6), team ($n$=1), organizational ($n$=1), and societal ($n$=1) levels. Specifically, LLMs assist software developers to sustaining cognitive flow, enhancing mental models, and offering a psychologically safe  and continuous learning environment. LLMs can also reduce developer effort and save developer time.
\end{tcolorbox}

\subsection{RQ2. How do LLMs create an imbalance in software development?} \label{subsec:disadvantages}

\subsubsection{Individual Level}

\textbf{Frictions in LLM-assisted software development.} Most of the participants (i.e., P1-P3, P5-P19, P21, P22) mention how using LLMs may slow software developers. For instance, they may need to start a new conversation when LLMs get stuck in the same wrong suggestion, e.g.: \faComments \;\textit{``I usually started from scratch and gave him as much background and information as possible. Because when I continued the same conversation, it tended to continue the same mistake"} -- P6 [Data Engineer].

\begin{table*}
\centering
\caption{Summary of the Challenges of using LLMs at the Individual (Software Practitioner) Level.}
\footnotesize
\begin{tblr}{
  width = \linewidth,
  colspec = {Q[86]Q[304]Q[344]},
  column{1} = {r},
  cell{1}{1} = {c},
  cell{2}{1} = {r=2}{},
  cell{4}{1} = {r=4}{},
  cell{8}{1} = {r=2}{},
  vlines,
  hline{1,10} = {-}{0.08em},
  hline{2,4,8} = {-}{},
  hline{3,5-7,9} = {2-3}{},
}
\textbf{\textbf{SE Activity}}                        & \textbf{\textbf{Software Development Tasks / How LLMs Impact?}}                                                                                                                                                                                                                                               & \textbf{Why the Impact Happens?}                                                                                                                                                                                                                                                                                                \\
{Requirement Engineering \& ~\\  Software Design}       & {\textbf{\textit{Task}}\textit{:}~Brainstorming\\\textbf{\textit{Challenge}}\textit{:}~[\#1] Risk of losing learning opportunities, [\#2] Hindering developers' skills}                                                  & \textbf{\textit{Reason:~}}Unpracticed skills$\text{}^{\#1,\#2}$; Handing over the control to LLM$\text{}^{\#1}$; Influence of personality$\text{}^{\#1,\#2}$; Overreliance on LLMs$\text{}^{\#1,\#2}$; Losing control over LLMs$\text{}^{\#2}$; LLMs reducing effort$\text{}^{\#1,\#2}$                                                                                                         \\
                                                     & {\textit{\textbf{\textbf{Tasks}}:}~Diagram Generation\\\textbf{\textit{Challenge: }}{[}\#1] Hindering developers' skills}                                                                                                                                 & \textbf{\textit{Reason:~}}Overreliance on LLMs$\text{}^{\#1}$; Losing control over LLMs$\text{}^{\#1}$; LLMs reducing effort$\text{}^{\#1}$; Mitigating interruptions$\text{}^{\#1}$; Unpracticed skills$\text{}^{\#1}$                                                                                                                                                                                       \\
{Software Development \& ~\\Software Quality Assurance} & {\textit{\textbf{\textbf{Tasks}}:}~Information Retrieval\\\textbf{\textit{Challenge: }}[\#1] Risk of losing learning opportunities, [\#2]~Hindering developers' skills}                                                                                        & \textbf{\textit{Reason:~}}Unpracticed skills$\text{}^{\#1}$; Handing over the control to LLM$\text{}^{\#1}$; LLMs reducing effort$\text{}^{\#1,\#2}$; Overreliance on LLMs$\text{}^{\#1,\#2}$; Influence of personality$\text{}^{\#1}$; Losing control over LLMs$\text{}^{\#2}$                                                                                                                                                     \\
                                                     & {\textit{\textbf{\textbf{Tasks}}:}~Code Understanding\\\textbf{\textit{Challenge: }}{[}\#1]~Hindering developers' skills}                                                                                                                              & \textbf{\textit{Reason:~}}Overreliance on LLMs$\text{}^{\#1}$; Losing control over LLMs$\text{}^{\#1}$; LLMs reducing effort$\text{}^{\#1}$                                                                                                                                                                                                                                     \\
                                                     & {\textit{\textbf{Task: }}Code Generation\\\textit{\textbf{Challenge: }}{[}\#1] Fragmented developer code mental model, [\#2] Hindering developers' skills, [\#3] Reputational vulnerability through LLM delegation, [\#4] Disrupt Developers' flow, [\#5] LLM-induced code quality risks} & \textbf{\textit{Reason:~}}Losing control over LLMs$\text{}^{\#1,\#2,\#3,\#4, \#5}$; Automation$\text{}^{\#1,\#4}$; Overreliance on LLMs$\text{}^{\#1,\#3,\#4}$; LLMs reducing effort$\text{}^{\#2,\#3}$; Mitigating interruptions$\text{}^{\#2}$; Unstable accuracy$\text{}^{\#3,\#5}$                                                                                                                                                  \\
                                                     & {\textit{\textbf{\textbf{Tasks}}:}~Test Generation; Test Data Generation; Code Translation; Pull Request Generation\\\textbf{\textit{Challenge: }}{[}\#1] LLM-induced code quality risks,~~[\#2] Hindering developers' skills}                                      & \textbf{\textit{Reason:~}\textit{}}Reliance on LLMs$\text{}^{\#1}$; Unstable accuracy$\text{}^{\#1}$; Unpracticed skills$\text{}^{\#2}$; Handing over the control to LLM$\text{}^{\#2}$; LLMs reducing effort$\text{}^{\#2}$; Overreliance on LLMs$\text{}^{\#2}$;  Mitigating interruptions$\text{}^{\#2}$; Influence of Expertise$\text{}^{\#2}$; Influence of personality$\text{}^{\#2}$; Losing control over LLMs$\text{}^{\#1}$ \\
Software Maintenance                                 & {\textit{\textbf{\textbf{Tasks}}:}~Debugging\\\textbf{\textbf{\textit{Challenge:~}}}~[\#1] Hindering developers' skills, [\#2] Overhead Introduced by LLMs}                                                                                   & \textit{\textbf{Reason:}} Unpracticed skills$\text{}^{\#1}$; Handing over the control to LLM$\text{}^{\#1}$; LLMs reducing effort$\text{}^{\#1}$; Overreliance on LLMs$\text{}^{\#1}$; Influence of personality$\text{}^{\#1}$; Mitigating interruptions$\text{}^{\#1}$; Lack of background knowledge$\text{}^{\#2}$; Unstable accuracy$\text{}^{\#2}$                                                                                                                                                     \\
                                                     & {\textit{\textbf{\textbf{Tasks}}:}~Code Review\\\textit{\textbf{Challenge: }}{[}\#1] LLM-induced code quality risks, [\#2] Losing learning opportunity}                                                                                                             & \textbf{\textit{Reason:}} Unstable accuracy$\text{}^{\#1}$; Losing control over LLMs$\text{}^{\#1}$; Handing over the control to LLM$\text{}^{\#2}$; LLMs reducing effort$\text{}^{\#2}$; Reliance on LLMs$\text{}^{\#2}$; Influence of personality$\text{}^{\#2}$; Mitigating interruptions$\text{}^{\#2}$                                                                                                                               
\end{tblr}
\label{tab:disadvantageIndividualLevel}
\end{table*}

\textbf{Frictions in human-LLM communication.} We identified six participants (i.e., P1-P4, P6, P17) mentioning challenges related to miscommunication with LLMs. This appears to happen due to AI hallucination or ``misunderstanding" of user prompts, e.g.: \faComments \;\textit{``Sometimes it will do extra things that I didn't ask for."} -- P5 [Software Developer]. But this can also occur due to the user facing difficulty in creating clear prompts, e.g.: \faComments \;\textit{``There's a communication problem. I can't explain the whole robust part so that the LLM has the basis to be able to answer me"} -- P6 [Data Engineer].

\textbf{Overhead introduced by LLMs.} The majority of the participants (i.e., P1, P3-P12, P14-P17, P19, P21, P22) presented scenarios where adopting LLMs increase their effort, for example, due to LLMs' response size limitation, e.g.: \faComments \;\textit{``The biggest [limitation] would be the question of answer size, where it can't answer a very large answer, and I'm going to have to ask several small questions in order to get my objective."} -- P1 [Software Developer]. In order to prevent LLMs from giving a full picture, software developers may also need to break down prompts into subprompts, e.g.: \faComments \;\textit{``I don't typically give the whole idea that I want to achieve to the GPT. Rather, I kind of break it down into different segments and then let them generate the code for me for each segment. In the end, I would do my own construction [, integration,] of these segments."} -- P17 [Data Analyst]. The unstable accuracy of LLMs also makes the developers spend energy in vain implementing wrong suggestions, e.g.:  \faComments \;\textit{``I would follow the debugging steps that they laid out for me. And then sometimes that works. Sometimes it doesn't, as usual"} -- P3 [Software Engineer].

\textbf{Situational LLM task inefficiencies.} Many participants  (i.e., P1, P3, P5-P8, P10-P11, P13-P17) reported situations where LLMs elongated rather than accelerated tasks. For instance, participants argue that there are situations where it is faster to do it themselves, e.g.: \faComments \;\textit{``Sometimes you have [moments when you get doubtful about LLMs]: `Okay, I'm just struggling to get this done because [LLMs] it doesn't get what I'm saying. So I'll just do myself'"} -- P10 [Software Engineer].

\textbf{LLM-induced code quality risks.} Most participants (i.e., P1-P3, P5, P7-P11, P13-P22) reported risks to code quality arising from LLM-assisted development. 
These risks manifest in three ways: first, over-reliance on LLMs may weaken developers' own skills, e.g.: \faComments \;\textit{``If you fully rely on it, the quality of your code base would drop by a lot"} -- P3 [Software Engineer]; second, LLMs often lack the organisational and project context needed to produce appropriate solutions, e.g.: \faComments \;\textit{``Usually I make a lot of changes to the code as well, because there's always something [that] the LLM will miss. The LLM won't have the entire context of the task, or [understand] what's the future plan [adopted by] the company"} -- P10 [Software Engineer]; third, LLMs may produce incorrect or incompatible suggestions, e.g.: \faComments \;\textit{``Sometimes ChatGPT - I don't know if it's [related to ChatGPT] design or something [else] - [recommend] libraries or packages [which] are not compatible. What ChatGPT proposed is not compatible with the environment I work in. I've encountered numerous times that I have to deal with this issue, and [I] ended up wasting a lot of time."} -- P17 [Data Analyst].

\textbf{Risk of losing learning opportunities.} Fourteen participants (i.e., P2-P8, P14-P15, P17, P19-P22) mentioned the potential negative impact on learning due to adopting LLMs. For instance, software developers might lose interesting discussions from online forums due to adopting LLMs, e.g.: \faComments \;\textit{``from Stack Overflow, [there are] more comprehensive discussions there"} -- P8 [Software Engineer]. Relying on LLMs to reduce effort via automation may take away valuable learning opportunities.

\textbf{Fragmented code mental models.} A few participants (i.e., P6, P10, P13-P14, P20-P21) mentioned the negative impact on the developer mental models.  They argue that when developers overrely on LLMs by handing over the control to them, they lose in code understanding. This also affects software developers' debugging capabilities, e.g.: \faComments \;\textit{``For example, Copilot developed five class methods for me, I didn't do it myself. If you debug it or show it to someone, you won't know how the code was implemented."} -- P6 [Data Engineer]. Consequently, they are pressured to continue relying on LLMs for debugging, e.g.: \faComments \;\textit{``If you have zero knowledge of the code, you just wrote because ChatGPT wrote it, and you didn't actually go through and read it, [doing code review], you then have to [rely on ChatGPT and to] ask ChatGPT to read the code that it wrote and then [get ChatGPT] tell it what problem happened"} -- P13 [Software Developer]. On the other hand, their code understanding improves naturally when coding by themselves, e.g.: \faComments \;\textit{``When you code yourself, you already have that [mental model] by default, because you had to code."} -- P10 [Software Engineer]. \label{subsubsec:incompleteModel}

\textbf{Erosion of developer skills and agency.} Sixteen participants (i.e., P1-P5, P7, P9, P11-P15, P19-P22) mention how adopting LLMs affects software developers' skills. This occurs because, when novice software developers overrely on LLMs to reduce effort by delegating tasks, they miss essential opportunities for the development of their technical and soft skills, e.g.: \faComments \;\textit{``You're so reliant on ChatGPT to think for you [...] your skills don't really increase much. You're still that beginner"} -- P5 [Software Developer]. For more experienced developers, the process of atrophying their software development skills happens due to reliance on LLMs, e.g.: \faComments \;\textit{``One of my colleagues said he deliberately turned off the [AI-based] completion because now he says `sometimes I just forget how it works. Sometimes I forget about the basics of my programming, because I'm used to this now'."} -- P12 [Software Engineer], \faComments \;\textit{``But since you're missing that sort of mental exercise, you don't really develop the coding muscles. And I think it kind of deteriorates your skills and your ability as a software developer."} -- P5 [Software Developer]. Doing code review is not defended as not being enough to stop the skill deterioration, e.g.: \faComments \;\textit{```Ah but [when using LLMs] you analyse the code', I do. But I think that even my capacity for analysis, it's lost over time. I'll spend more and more time just accepting, just like [when] you review a PR (pull request), and you don't really care. I think I'd do the same thing over time, gradually [...] my review would get worse."} -- P4. LLMs' automation capabilities influence software developers in becoming lazier, e.g.: \faComments \;\textit{``I have become lazy [to write] even if [it is just a] small one line I can also rely on the LLMs [...]  if I have to read a documentation, I will just ask LLMs to summarise so that I don't have to read it completely"} -- P19 [Machine Learning Scientist], apathetic, e.g.: \faComments \;\textit{``Even my capacity for analysis has been lost over time. I'll spend more and more time just accepting, just like you review a PR (pull request). And you don't really care"} -- P4 [Software Developer], and less confident in their own development skills, e.g.: \faComments \;\textit{``It would take much longer to do something [that while not using AI, compared to] you [that] can do it with ChatGPT or Claude nowadays."} -- P15 [Researcher]. \label{subsubsec:skillAtrophy}

\textbf{Reputational vulnerability through LLM delegation.} A few participants (i.e., P6-P7, P9-P10, P19, P21-P22) mention how errors from code generated via LLMs can affect developers' credibility, due to LLMs' non-deterministic nature. For this reason, participants demonstrate a cautious approach towards completely delegating tasks to LLMs, e.g.: \faComments \;\textit{``It [code with errors] impacts your credibility as a professional"} -- P9 [AI Engineer], \faComments \;\textit{``When you push the code to the Git [repository], it is not the LLM's name that will be in that log, it will be your name."} -- P19 [Machine Learning Scientist].

\label{subsubsec:disruptflow} 
\textbf{Developer flow interruptions through LLM interaction.} We found eleven participants (i.e., P4-P5, P7-P8, P12-P13, P15-P19) mentioned how interaction with LLMs can disturb their flow. For instance, interacting with LLMs via their online platform may cost a context switch to the code editor, e.g.: \faComments \;\textit{``If I need to find a solution with it, I'm still [do] context switch because [ChatGPT] it's not inside [the IDE] where I'm coding"} -- P16 [R\&D Researcher]. The many suggestions offered by LLMs can also distract software developers, e.g.: \faComments \;\textit{``Sometimes ChatGPT will give you an answer that's completely different to what you were thinking. Unless you actually implement those [potential solutions] in real life, you don't know what the result is going to be [...] [basically] ChatGPT puts another option on top of your solution that you have in your mind."} -- P7 [BI Analyst].

\subsubsection{Team Level}

\textbf{Losing mentorship opportunities.} Some participants (i.e., P2-P3, P6-P7, P15-P14, P18-P19, P21-P22) describe how LLMs negatively influence mentorship opportunities, taking the mentorship role from senior developers. This happens because novice developers would seek assistance first from LLMs, \faComments \;\textit{``If I'm researching something that I don't know a lot about, I usually will turn to technical documentation and Google [...] first. While mentoring a junior in my field, they were very quick to go to LLMs first, and even criticise me not for going to Google first"} -- P22 [Software Developer].

\subsubsection{Organisation Level}

\textbf{Issues with the code license.} A few participants (i.e., P2, P11, P16) mention license concerns at the organisation level, and how they shape their organisation's approach towards LLMs. Seeking to avoid potential code license issues, companies can decide to prohibit using LLMs in the workplace, e.g.: \faComments \;\textit{``At my previous work [...] their policy was just to avoid using ChatGPT or any other LLMs. Their concern was more to do with the ownership of the code that was being generated, and how it might violate the trademarks."} -- P2 [Software Engineer]. Settling licenses may also cause a delay in the adoption of LLMs at the workplace, e.g.: \faComments \;\textit{``We had to wait for the licenses [regarding ChatGPT] to come for that"} -- P16 [R\&D Researcher].

\textbf{Issues with security \& privacy.} We found security and privacy concerns in the responses of twelve participants (i.e., P2-P3, P5, P7, P9-P11, P13-P16, P18). There is a sceptical attitude on how LLMs approach these topics, e.g.: \faComments \;\textit{``For security reasons, my current company isn't allowing AI [to be] integrated into the program."} -- P7 [Data Analyst].

\textbf{Cost of using LLMs.} Seven participants (i.e., P2, P9-P10, P12, P16-P17, P19) mentioned the organisational cost of adopting LLMs. In this sense, the cost of tokens may limit high-volume interactions (i.e., \textsl{vibe coding}), e.g.: \faComments \;\textit{``When we try to scale [that AI-generated code], we know that it becomes much faster and much more expensive quickly because we rely on OpenAI to do the calls, and we pay by token. You can't actually do it because it's going to cost this amount.} -- P16 [R\&D Researcher].

\subsubsection{Societal Level}

\textbf{Erosion of social trust.} We found negative effects of LLMs on social trust in six participants (i.e., P2, P11, P14, P16, P19, P22). There is a concern of LLMs being misused, such as cheating during interview processes, e.g.:  \faComments \;\textit{``There's also a lot of LLMs that can help you pass an interview, right? [Of course] it really depends on how [the interview] is monitored."} -- P14 [Data Engineer], \faComments \;\textit{``People would still need to consider the situation where it would add to misinformation and generate maybe a lot of fake news and fake videos [...] [because of] that people would find it difficult to trust anything"} -- P2 [Software Engineer].

\textbf{Job market crisis.} Eight participants (P3, P7, P12, P16, P18, P20-P22) describe the impact of LLMs on the job market. Automation, not just LLMs, appears as the motivation for human replacement in simple tasks, e.g.: \faComments \;\textit{``With the rise of AI and automation, a lot of things that humans can do can be replaced by using a robot."} -- P3 [Software Engineer]. While there is still a demand for experts, non-specialised professionals might be replaced,  e.g.: \faComments \;\textit{``We're still gonna need to rely on those experts for a lot of things. But then the issue is the ones that are not the experts, the in-between ones."} -- P16 [R\&D Researcher].

\begin{tcolorbox}[float=htpb!,colback=black!5!white,colframe=black!75!black,title={\textbf{RQ2.} \textit{How do LLMs create an imbalance in software development?}}]
Our findings reveal challenges distributed across the individual ($n=10$), team ($n=1$), organizational ($n=3$), and societal ($n=2$) levels. Specifically, the adoption of LLMs in software development tasks can introduce operational friction, increase cognitive overhead, and compromise code quality. Furthermore, LLM reliance may lead to the erosion of developer skills, codebase familiarity, mental flow, and professional reputation.
\end{tcolorbox}

\subsection{RQ3. How can software developers achieve a balanced use of LLMs?} \label{subsec:recommendations}

Our findings on the challenges of adopting LLMs for software development, as presented in Section \ref{subsec:disadvantages}, drive us to recommend actionable guidance to address these drawbacks based on our interview findings. We present the recommendations followed by the addressed challenges.

\noindent \textbf{\faHandORight \; Visualise code using AI tools (e.g., sequence diagram, UML diagram)}. This helps developers to understand and assess how the code works, e.g.: \faComments \;\textit{``So if I have described an architecture. [I can say to the LLMs:] `Create the diagram for this architecture'. [Then] it creates the mermaid diagram, so it can easily be visualised how this architecture works."} -- P10 [Software Engineer]. 

\begin{itemize}
    \item \textit{Challenge}: Fragmented code mental model.\\\textit{Rationale}: The visualisation allows developers to create a mental model of code generated using LLMs. 
     \item \textit{Challenge}: LLM-induced code quality risks.\\\textit{Rationale}: The visualisation allows developers to review the workflow used in the code generated via LLMs.
     \item \textit{Challenge}: Risk of losing learning opportunities.\\\textit{Rationale}: LLMs can suggest more appropriate architectures, overlooked by developers. This allows developers to improve their understanding of software design.
\end{itemize}

\noindent \textbf{\faHandORight \; Adopt self-hosted LLM solutions.} Participants P9 and P22 run LLMs (i.e., Jan.AI, Llama, H2O Danube, and Mistral) locally via Ollama, Llama.cpp, and LM Studio as an approach to deal with privacy concerns in hosted LLM services. 

\begin{itemize}
    \item \textit{Challenge}: Issues with security \& privacy.\\\textit{Rationale}: Employing self-hosted solutions eliminates concerns regarding the organisation's data privacy.
    \item \textit{Challenge}: Risk of losing learning opportunities.\\\textit{Rationale}: Deploying self-hosted solutions allows developers to have a better understanding of LLMs' limitations.
\end{itemize}

\noindent \textbf{\faHandORight \; Integrate solutions, not just LLMs, into the workflow.} Participant P9 defends that Copilot is an example of an AI-integrated solution, which comprehends GitHub Actions and GitHub -- popular continuous integration and continuous delivery and code versioning platforms. Participant P22 incorporated an AI-assisted code review by connecting Ollama into their GitLab server.

\begin{itemize}
    \item \textit{Challenge}: Developer flow interruptions through LLM interaction.\\\textit{Rationale}: Employing integrated LLM solutions allows for more contextualised and accurate assistance, reducing the need for human intervention.
\end{itemize}

\noindent \textbf{\faHandORight \; Combine different smaller models.} After experimenting with H2O.ai\footnote{\url{https://h2o.ai}}, Participant P9 suggests using Small Language Models as a solid alternative to LLMs, e.g.: \faComments \;\textit{``In some parts of the solution, it's possible to use smaller models very efficiently and diminish the cost of the overall solution"} -- P9 [AI Engineer].  

\begin{itemize}
    \item \textit{Challenge}: Cost of using LLMs.\\\textit{Rationale}: Employing smaller language models rather than large ones can help reduce computational, memory, and energy requirements \cite{argerich:2024}.
\end{itemize}

\noindent  \textbf{\faHandORight \; Prohibit LLMs from merging code into the main branch.} Human in the loop is still required for accountability, e.g.: \faComments \;\textit{``If I’m putting up a pull request for my company, I need to understand that code. I need to have the ownership of it. Because if something fails, I won’t be blaming Cursor, or anyone else, this [error] will be on me to fix. Any code that’s generated by the LLMs, I only put this code in production [environment] if I actually understand that [code]"} – P10 [Software Engineer].

\begin{itemize}
     \item \textit{Challenge}: LLM-induced code quality risks.\\\textit{Rationale}: Implementing mandatory human review for main-branch commits can safeguard against errors in LLM-generated code \cite{sun:2025}.
     \item \textit{Challenge}: Reputational vulnerability through LLM delegation.\\\textit{Rationale}: The credibility of developers and their organisations is preserved by safeguarding against errors from code generated via LLMs. 
\end{itemize}

\noindent  \textbf{\faHandORight \; Follow the established coding conventions.} Participant P11 described how GitHub Copilot can get confused according to the used variables, e.g.:  \faComments \;\textit{``When you have some kind of misleading or not super standard variable names, Copilot gets confused"} -- P11 [Web Developer]. 

\begin{itemize}
    \item 
    \textit{Challenge}: Frictions in human-LLM communication.\\\textit{Rationale}: By adhering to code conventions consistent with those found in the data used to train LLMs, developers can reduce the risk of miscommunication with these models.
    \item \textit{Challenge}: LLM-induced code quality risks.\\\textit{Rationale}: By adhering to code convention best practices, code generated by LLMs is more likely to achieve satisfactory quality.
    \item \textit{Challenge}: Reputational vulnerability through LLM delegation.\\\textit{Rationale}: By mitigating errors from code generated using LLMs, the credibility of developers and their organisations is preserved.
    \item \textit{Challenge}: Erosion of developer skills and agency.\\\textit{Rationale}: By following code conventions, developers.
\end{itemize}

\noindent \textbf{\faHandORight \; Introduce LLM4SE usage policies.} Four participants (18.1\%) indicated in the pre-interview questionnaire that their organisations either do not have an AI policy or are not aware of one. Having a clear policy addressing the use of software development forms an essential foundation for effective IT governance.

\begin{itemize}
    \item \textit{Challenge}: Issues with security \& privacy.\\\textit{Rationale}: LLMs4SE policies that indicate the supported LLMs tools and how to share code with those tools may mitigate concerns regarding privacy.
    \item \textit{Challenge}: Issues with code license.\\\textit{Rationale}: Defining how to handle code licensing in the LLM4SE policy can guide developers and help prevent potential legal issues in the future.
    \item \textit{Challenge}: Erosion of social trust.\\\textit{Rationale}: Clarifying how LLMs manage developers' prompts may help ease their concerns.
\end{itemize}

\noindent \textbf{\faHandORight \; Continuously upskilling in the latest SE practices and tools.} Participants P1, P4, P10, P14, and P20 defend the importance of upskilling to not get behind in the market, e.g.:  \faComments \;\textit{``I will always try to upskill myself in a way that I won't get replaced by LLMs because the implementations are not perfect"} -- P20 [Software Engineer].

\begin{itemize}
    \item \textit{Challenge}: Job market crisis.\\\textit{Rationale}: Keeping up-to-date with the latest SE best practices and tools, such as spec-driven development, may help developers to maintain relevance in this AI era.
\end{itemize}

\section{Discussion}
\label{sec:discussion}

\subsection{Implications for Practice}\label{implications4practice}

We analyse the implications of adopting LLMs in SE by exploring both the benefits (RQ1) and challenges (RQ2) using Marshall McLuhan’s Tetrad of Media Effects \cite{mcluhan:1977}. This framework enables us to examine how new technologies can augment human capabilities through different aspects. The Tetrad consists of four fundamental questions: \textit{What does the technology \textbf{enhance} or intensify? What does it make \textbf{obsolete} or displaced? What does it \textbf{retrieve} or recover from the past? And what does it \textbf{reverse} into when taken to extremes?}

\textsc{Enhance.} This explores how adopting LLMs enhances or intensifies practices and processes in software development.

\begin{itemize}

    \item \textbf{Cognitive debt}: In section \ref{subsubsec:incompleteModel}, we highlight how adopting LLMs can deprive developers of a complete mental model of the code. According to Storey \cite{storey:2026}, this cognitive debt may silently build up over time and lead to developers losing control over their software. In our study, participants keep control over LLMs by (i) relying on them for code improvement rather than code generation and (ii) employing Test-Driven Development (TDD). Mathews et al. \cite{mathews:2024} found that employing TDD with models like GPT-4 and Llama 3 leads to a higher success rate in solving programming challenges.
    
    \item \textbf{Company onboarding}: In section \ref{subsubsec:onboarding}, we present the potential of LLMs being incorporated into the onboarding process. However, in the ongoing case study by Azanza et al. \cite{azanza:2024} at an IT consulting organisation, they found that using third-party LLMs can pose unacceptable data risks for companies. They explored open-source LLMs for onboarding. Additionally, organisations may seek agreement with LLM vendors for data protection.

\end{itemize}

\textsc{Obsolete.} This aspect explores the obsolescence or displacement caused by adopting LLMs for software development.

\begin{itemize}
    \item \textbf{Repetitive tasks}: In section \ref{subsubsec:repetitiveTasks}, we mention how LLMs reduce developers' effort for simple tasks, such as boilerplate code generation and information retrieval. However, it is still essential to keep humans in the loop, at least as reviewers, since LLMs are based on stochastic models.
\end{itemize}

\textsc{Retrieve.} This examines how far adopting LLMs makes it possible to retrieve or recover concepts, practices, or processes adopted in the past in SE.

\begin{itemize}
    \item \textbf{Exploratory learning}: In section \ref{subsubsec:exploratoryLearning}, we highlight how participants use LLMs as a learning tool, augmenting the traditional problem-solving process. Using the chat mode, similar to Hallway talks, developers can ask questions, simulating hypothetical scenarios, getting an independent perspective. We suggest that other developers be added to the loop for sharing knowledge purposes. 

\end{itemize}

\textsc{Reverse.} This explains how adopting LLMs for software development, when pushed to its extreme, undergoes a reversal or transformation.
\begin{itemize}
    \item \textbf{Skill atrophy}: In section \ref{subsubsec:skillAtrophy}, we highlighted that adopting LLMs may hinder developers' skills. While LLMs can provide time-saving for software development by reducing effort, that effort is also necessary to practice their skills. In our study, participants restrict themselves from overusing AI (e.g., turning off AI-based suggestions once in a while) and force themselves to use their coding muscles.

    \item \textbf{Environmental crisis}: When it comes to environmental cost, surprisingly, only two participants (P16, P22) demonstrated concerns about energy consumption, overlooking water consumption \cite{jegham:2025}. Shi et al. \cite{shi:2025} suggested that developers employ program-centric techniques, such as program pruning and grammar augmentation, to reduce the number of tokens.

\end{itemize}

\subsection{Implication for Research} \label{sec:implicationsToResearch}

Based on the findings, we identified research gaps related to software practitioners adopting LLMs for Software Development.

\textbf{What factors influence developers' decisions to delegate SE tasks to LLMs?} Our interview participants described adopting different attitudes regarding the use of LLMs for SE tasks. We observed that task risk and developers' motivation play an undeniable role in task delegation to LLMs. For instance, Participant P19 argues against relying on LLMs for test generation, warning that it could create a false sense of security.  Regarding developers' motivation, a lack of enthusiasm for certain tasks (e.g., updating documentation) among developers seems to encourage them to delegate such responsibilities. We suggest further studies to investigate other factors and how they interplay with task risk and developers' motivation.

\textbf{How do LLMs empower developers during tight deadlines?} 
Our data reveals a spectrum of behaviours: whereas some developers use LLMs to meet tight deadlines rapidly, others may inflate their effort estimations when motivation is low. Researchers should further examine the triggers for such inflation to better understand the long-term impact of AI assistance on software process metrics.


\textbf{How do LLMs affect social intelligence?} Our findings show that using LLMs may mitigate interruptions at the team level. Although it may benefit in terms of productivity, the software development life cycle is entwined with \textit{``collaboration, human judgment, emotional interactions, and decision-making"} \cite{alami2:2025, kalliamvakou:2017}. This is particularly important for manager positions, where social intelligence plays a crucial role in capitalising on interactions with customers and employers.

\textbf{How do LLMs affect developers' intuition?} According to Naur \cite{naur:1985}, \textit{``Software development in all its phases, and irrespective of the techniques employed in its pursuits, must and will always depend on intuition"}. Intuition built from previous experiences acts as a safeguard mechanism, enabling software developers to navigate LLMs' wrong suggestions. While Chen et al. \cite{chen:2023} investigated the role of human intuition when using LLMs, the study regarding the effects of LLMs on human intuition remains unexplored.


\textbf{How does agentic software engineering contribute to developers' burnout?} Investigations into burnout in SE have been the focus of numerous studies \cite{tulili:2023}.  In their systematic mapping study, Tulili et al. \cite{tulili:2023} identified the following causes for burnout: personality traits (e.g., neuroticism), work-related factors (e.g., job demand, job overload), communication practices, agile practices, and physiological factors. However, software developers' burnout from adopting LLM tools is an emergent topic that requires attention from researchers. A recent large survey conducted by Feng et al. \cite{feng:2025} found that job demands—such as organisational pressures and increased workload—are positively associated with burnout, while job resources related to generative AI adoption—such as autonomy and access to learning resources—are negatively associated with it. Beyond the adoption of generative AI, the rise of agentic software development can place a considerable burden on developers, who may need to supervise several agentic tools doing different software engineering tasks in parallel. Future research should investigate how this expanding supervisory burden interacts with established burnout drivers — and critically, what constitutes a sustainable threshold of concurrent agent supervision for software developers.    

\textbf{How does agentic software engineering influence job title inflation within the tech industry?} Title inflation occurs when employees are granted senior-level designations that exceed their actual duties or compensation \cite{rampton:2026}. In the era of agentic workflows, developers across all seniority levels are increasingly tasked with overseeing autonomous agents in addition to their traditional responsibilities. However, as these roles evolve to include agent orchestration, compensation structures often remain static. We propose that future studies investigate whether this gap between responsibility and reward is being bridged by inflated job titles and what impact this has on the long-term transparency of the software labour market.

\textbf{What are the most appropriate LLM maturity model stages for organisations?} Our data shows varying degrees of LLM adoption, ranging from superficial tool access (e.g., GitHub Copilot for code completion) to deep integration within DevOps pipelines and code review processes. While the Quotient maturity framework \cite{quotient:2026} identifies five stages—ranging from ad hoc adoption to end-to-end autonomy—and suggests that stages 3 (standardised workflows) and 4 (supervised automation) are the most appropriate targets, these recommendations lack broad empirical support. We advocate for longitudinal studies to validate which maturity levels actually yield the highest return on investment and team stability in practice.

\textbf{Controversial use cases.} Participants demonstrate different opinions regarding using LLMs for test generation, code generation, and debugging. LLMs are discouraged for test-related tasks due to providing the illusion of false assurance. Wang et al. \cite{wang:2024} argue that using LLMs for integration tests may result in errors or unreliable results because those tests may exceed the capacity of the LLM to process and analyse. LLMs are also discouraged for code generation due to the harmful long-term effects towards software developers' skills. Some participants also highlight that LLMs may provide general debugging guidance, since they cannot run the code.

\section{Related Work}\label{sec:relatedWork}

\begin{figure*}
    \centering
    \includegraphics[width=0.85\linewidth]{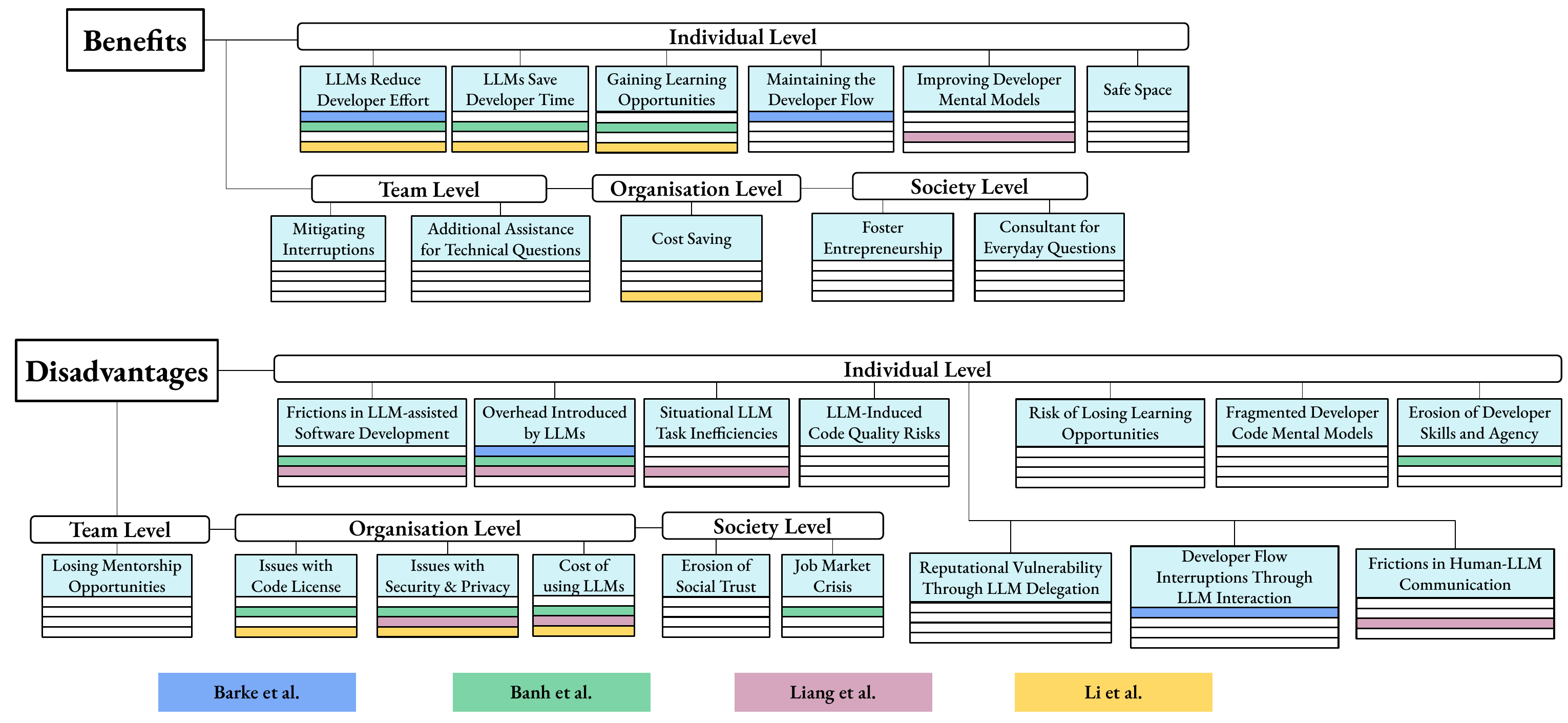}
    \caption{Comparison with related work focused on benefits and challenges. The colored bars represent the overlap with existing literature \cite{barke:2023, banh2:2025, liang:2025, li:2024}, and empty bars indicate new findings.}
    \label{fig:ComparisonRelateWork}
\end{figure*}

Barke et al. \cite{barke:2023} conducted an observational and interview study with twenty software developers using GitHub Copilot on how they interact with GitHub Copilot.  They employed traditional Grounded Theory analysis \cite{glaser:2017}, identifying two interaction modes: acceleration mode and exploration mode. In acceleration mode, the developer has a clear understanding of the next required task; whereas developers explore GitHub Copilot autocomplete suggestions in exploration mode due to a lack of direction. According to Barke et al., the interactions in the acceleration mode are fast and do not disrupt the programmer's flow. Our study findings corroborate their findings regarding LLMs maintaining development flow. Also aligned with our findings, they identified that LLMs can also disrupt the development flow when providing long suggestions.

Banh et al. \cite{banh2:2025} conducted eighteen interviews with IT-related professionals (i.e., software developers, a product owner, and a scrum master) between August 2023 and January 2024. They aimed to understand how to integrate Generative AI in Software Engineering, exploring the opportunities and challenges related to adopting Generative AI tools. They applied the traditional Grounded Theory, resulting in the conceptual framework of generative AI integration in Software Engineering practices. Their findings, similar to ours, are: reduced development time, issues with intellectual property, underestimated overhead, developer empowerment, and code quality improvement via LLMs. However, our findings also include the contradictions involving the benefits and challenges, such as code degradation via LLMs, LLMs dragging tasks (see \ref{subsubsec:disruptflow}). Our results point out the necessity for a balanced approach to the adoption of LLMs, advocating for their use as a net positive in the software development process. Our recommendations in Section \ref{subsec:recommendations} contain practical suggestions contributing to achieve this.

Liang et al. \cite{liang:2025} employed traditional Grounded Theory \cite{juliet:2015} through twenty interviews and observation of fifteen software developers. We could not identify when interviews and observation were conducted. Their contribution includes an understanding of prompt programming practices. Similar to our study, they identified the potential of LLMs to assist developers in improving their mental model. Although their research focused on prompt programming aspects, they present their impact across different software development tasks similar to us (See Table \ref{tab:benefitsIndividualLevel} and \ref{tab:disadvantageIndividualLevel}). Our findings distinguish by showing not only the impact of software practitioners' interaction with LLMs, but also how it reflects in others levels, such as team level (e.g., mitigating interruptions).

Li et al., \cite{li:2024} conducted twenty-six interviews in three rounds with industry practitioners and 395 survey respondents. They employed socio-technical grounded theory \cite{hoda:2022, hoda:2024}, which resulted in the theory of AI tool use and adoption in software engineering. Similar to our study, they also classified the impact at the individual and organisational level, such as fear of decreased skills and potential judgment of using LLMs, which are related to the sections hindering developers' skills and damaging developers' reputation from our study (See \ref{subsec:disadvantages}). On the other hand, our findings also comprehend the impact at the team and society level, such as losing mentorship opportunities due to LLMs, and LLMs fostering entrepreneurship.

Although the literature presents studies exploring the impact of LLMs for software development, as illustrated in Figure \ref{fig:ComparisonRelateWork}, our investigation goes further by identifying best practices to balance the forward and backward impact of adopting LLMs.

\section{Evaluation}

\subsection{Evaluating STGT4DA Application}

The evaluation of the STGT4DA application consists of \textit{credibility} and \textit{rigour} as key criteria \cite{hoda:2022, hoda:2024}. In terms of \textbf{credibility}, our section \ref{sec:methodology} provides details involving the recruitment process (social media and emailing), sampling method (purposive sampling followed by convenient sampling), how iterative and interleaved data occurred (three rounds of data collection and analysis), and how memos were written and applied (to guide the structure of emerging concepts and sub-categories, and as directions for future work). Concerning \textbf{rigour}, our section \ref{sec:methodology} also provides examples of basic coding and constant comparison (See Fig. \ref{fig:OpenCodeConstantComparison}) and embedded sanitised evidence (i.e., several interview quotes throughout Section \ref{sec:findings}).

\subsection{Evaluating STGT4DA Outcomes}

Findings from the application of STGT4DA should exhibit \textit{originality}, \textit{relevance}, and \textit{density} \cite{hoda:2022, hoda:2024}. In terms of \textbf{originality}, we position our findings in comparison to related work in Section \ref{sec:relatedWork}. The \textbf{relevance} of understanding the impact of LLMs on software practitioners is highlighted in Sections \ref{sec:introduction}, \ref{implications4practice}, and \ref{sec:implicationsToResearch}. Regarding \textbf{density}, Section \ref{sec:findings} condenses the sub-categories and concepts that compose the category ``\textsl{Impact on using LLMs}". We illustrate the depth of the findings through multiple levels of codes, concepts, and sub-categories, and the richness of findings through various aspects of the phenomenon, illustrated through numerous interview quotes.

\section{Threats to Validity}
\label{sec:limitations}

We will discuss the study limitations using the Total Quality Framework (TQF) developed by \cite{roller:2015}, suitable for qualitative studies in software engineering \cite{lenberg:2024, gunatilake:2025}. This framework is structured in the following aspects: \textit{credibility} regarding data collection, \textit{analyzability} of the data, \textit{transparency} of reporting, and \textit{usefulness} of the findings.

\textbf{Credibility.} It refers to how comprehensively and accurately the \textit{data collection} was performed  \cite{korstjens:2018}. Our study participants cover individuals from different demographics, such as gender, country of residence, years of professional experience, and role. At the same time, we acknowledge that future research could expand the participants' pool by interviewing, for example, IT managers and project managers or collecting data from professional social media (e.g., LinkedIn and Reddit posts). Besides our challenges in recruiting participants - an arduous task described by \cite{madampe:2024} - we reached a suitable sampling size. We also acknowledge that changing the ATAI scale questions from 9 scale to 5 scale may impose threats to validity. However, we also defend that it may facilitate respondents to answer the questions. 

\textbf{Analyzability.} It refers to how comprehensively and accurately the \textit{data analysis} was designed and executed. Although our qualitative analysis carries an intrinsic subjectivity, we mitigate this by conducting a pilot for the open coding, as well as conducting discussions over the emerging codes, concepts, sub-categories and categories between the authors.

\textbf{Transparency.} It refers to how clear and complete this paper reports the aspects referring to credibility and analysability, supporting replication or transference to other contexts. We intend to present enough information. Additionally, we provide further information in the appendices and in the online supplementary material package \cite{researchArtifact}.

\textbf{Usefulness.} It refers to how useful the study's contributions and findings are. Our findings can support software team leaders and IT managers in evaluating whether LLMs align with their context and needs, as well as guiding SE researchers for future research needs. We acknowledge that the fast-paced evolution of LLMs can quickly make some of our findings outdated. However, empirical studies help capture and document the evolving state of practice, especially as previous models become inaccessible. Further investigations can compare to our findings and report on the latest LLM capabilities.

\section{Conclusion} \label{sec:conclusion}

In this paper, we present the software practitioners' perspective on the impact of LLMs at the individual, team, organisation, and society levels. Our findings comprehend benefits and challenges of using LLMs for software development. At the individual level, most of the participants mention LLMs saving time and reducing effort as benefits, and frictions and overhead introduced by LLM-assisted software development as challenges. At the team level, most of the participants refer to LLMs as mitigating interruptions, but also reducing mentorship opportunities. At the organisation level, participants highlight the cost savings due to LLMs, but also security and privacy concerns. At the society level, participants mention LLMs promoting entrepreneurship, but also the erosion of social trust. The challenges identified create an imbalance in the use of LLMs. To address this issue, our findings also include actionable suggestions to help mitigate this imbalance and support developers as they navigate the integration of LLMs into software development.

\section{Acknowledgment}

Samuel is supported by a Faculty of IT PhD scholarship. We would like to express profound gratitude to all participants who took part in this study, Hashini Gunatilake, Humphrey Obie, and Marcos Medeiros, for the wonderful feedback on this research, and Nimmi Weeraddana and Haoyu Gao for helping during participant recruitment.

\bibliographystyle{IEEEtran}
\bibliography{bibliography}

@String{Computing = "Computing" }

@String{Computer = "{IEEE} Computer" }

@String{Springer = "Springer-Verlag" }

@article{hoda:2022,
  title={Socio-technical grounded theory for software engineering},
  author={Hoda, Rashina},
  journal={IEEE Transactions on Software Engineering},
  volume={48},
  number={10},
  pages={3808--3832},
  year={2022},
  publisher={IEEE}
}

@book{hoda:2024,
    author = {Hoda, Rashina},
    title = {Qualitative Research with Socio-Technical Grounded Theory},
    publisher = {Springer},
    year = {2024},
    url = {https://link.springer.com/book/10.1007/978-3-031-60533-8}
}

@online{storey:2026,
  author = {Storey, Margaret-Anne},
  title = {How Generative and Agentic {AI} Shift Concern from Technical Debt to Cognitive Debt},
  year = {2026},
  url = {https://margaretstorey.com/blog/2026/02/09/cognitive-debt/},
  urldate = {2026-02-23}
}

@inproceedings{azanza:2024,
  title={Can {LLMs} facilitate onboarding software developers? an ongoing industrial case study},
  author={Azanza, Maider and Pereira, Juanan and Irastorza, Arantza and Galdos, Aritz},
  booktitle={2024 36th International Conference on Software Engineering Education and Training (CSEE\&T)},
  pages={1--6},
  year={2024},
  organization={IEEE}
}

@article{mcluhan:1977,
  title={Laws of the Media},
  author={McLuhan, Marshall},
  journal={ETC: A Review of General Semantics},
  pages={173--179},
  year={1977},
  publisher={JSTOR}
}

@article{ziegler:2024,
  title={Measuring GitHub Copilot's impact on productivity},
  author={Ziegler, Albert and others},
  journal={Communications of the ACM},
  volume={67},
  number={3},
  pages={54--63},
  year={2024},
  publisher={ACM New York, NY, USA}
}

@article{ebert:2023,
  title={Generative {AI} for software practitioners},
  author={Ebert, Christof and Louridas, Panos},
  journal={IEEE Software},
  volume={40},
  number={4},
  pages={30--38},
  year={2023},
  publisher={IEEE}
}

@article{liang:2025,
  title={Prompts are programs too! understanding how developers build software containing prompts},
  author={Liang, Jenny T and others},
  journal={Proceedings of the ACM on Software Engineering},
  volume={2},
  number={FSE},
  pages={1591--1614},
  year={2025},
  publisher={ACM New York, NY, USA}
}

@inproceedings{krauss:2025,
  title={"Create a Fear of Missing Out"-ChatGPT Implements Unsolicited Deceptive Designs in Generated Websites Without Warning},
  author={Krau{\ss}, Veronika and others},
  booktitle={Proceedings of the 2025 CHI Conference on Human Factors in Computing Systems},
  pages={1--20},
  year={2025}
}

@article{kuhail:2024,
  title={“Will I be replaced?” Assessing ChatGPT's effect on software development and programmer perceptions of {AI} tools},
  author={Kuhail, Mohammad Amin and  others},
  journal={Science of Computer Programming},
  volume={235},
  pages={103111},
  year={2024},
  publisher={Elsevier}
}

@book{juliet:2015,
  title={Basics of qualitative research: Techniques and procedures for developing grounded theory},
  author={Juliet, M and Corbin, Strauss},
  year={2015},
  publisher={SAGE Publications, Incorporated}
}

@book{glaser:2017,
  title={Discovery of grounded theory: Strategies for qualitative research},
  author={Glaser, Barney and Strauss, Anselm},
  year={2017},
  publisher={Routledge}
}

@article{chen:2023,
  title={Understanding the role of human intuition on reliance in human-AI decision-making with explanations},
  author={Chen, Valerie and others},
  journal={Proceedings of the ACM on Human-computer Interaction},
  volume={7},
  number={CSCW2},
  pages={1--32},
  year={2023},
  publisher={ACM New York, NY, USA}
}

@article{mayer:2025,
  title={Superagency in the workplace: Empowering people to unlock {AI}’s full potential},
  author={Mayer, Hannah and others},
  journal={McKinsey Digital},
  volume={28},
  year={2025}
}

@article{mohamed:2025,
  title={The Impact of LLM-Assistants on Software Developer Productivity: A Systematic Literature Review},
  author={Mohamed, Amr and Assi, Maram and Guizani, Mariam},
  journal={arXiv preprint arXiv:2507.03156},
  year={2025}
}

@inproceedings{naur:1985,
  title={Intuition in software development},
  author={Naur, Peter},
  booktitle={International Joint Conference on Theory and Practice of Software Development},
  pages={60--79},
  year={1985},
  organization={Springer}
}

@inproceedings{nam:2024,
  title={Using an llm to help with code understanding},
  author={Nam, Daye and Macvean, Andrew and Hellendoorn, Vincent and Vasilescu, Bogdan and Myers, Brad},
  booktitle={Proceedings of the IEEE/ACM 46th International Conference on Software Engineering},
  pages={1--13},
  year={2024}
}

@article{barke:2023,
  title={Grounded copilot: How programmers interact with code-generating models},
  author={Barke, Shraddha and others},
  journal={Proceedings of the ACM on Programming Languages},
  volume={7},
  number={OOPSLA1},
  pages={85--111},
  year={2023},
  publisher={ACM New York, NY, USA}
}

@online{dora:2024,
  title={{Superagency in the workplace: Empowering people to unlock {AI}'s full potential}},
  author={Debellis, Derek and others},
  url       = {https://dora.dev/research/ai/gen-ai-report/},
  year={2024},
  urldate   = {2025-09-26}
}

@online{dora:2025,
  title        = {{2025 DORA State of AI-Assisted Software Development}},
  author       = {DeBellis, Derek and others},
  year         = {2025},
  url          = {https://dora.dev/research/ai/#state-of-ai-assisted-software-development},

  urldate   = {2025-09-26}
}

@article{cui:2024,
  title={The effects of generative {AI} on high skilled work: Evidence from three field experiments with software developers},
  author={Cui, Zheyuan Kevin and Demirer, Mert and Jaffe, Sonia and Musolff, Leon and Peng, Sida and Salz, Tobias},
  journal={Available at SSRN 4945566},
  year={2024}
}

@article{li:2024,
  title={AI tool use and adoption in software development by individuals and organizations: a grounded theory study},
  author={Li, Ze Shi and others},
  journal={arXiv preprint arXiv:2406.17325},
  year={2024}
}

@article{banh2:2025,
  title={Copiloting the future: How generative {AI} transforms Software Engineering},
  author={Banh, Leonardo and Holldack, Florian and Strobel, Gero},
  journal={Information and Software Technology},
  volume={183},
  pages={107751},
  year={2025},
  publisher={Elsevier}
}

@article{korstjens:2018,
  title={Series: Practical guidance to qualitative research. Part 4: Trustworthiness and publishing},
  author={Korstjens, Irene and Moser, Albine},
  journal={European Journal of General Practice},
  volume={24},
  number={1},
  pages={120--124},
  year={2018},
  publisher={Taylor \& Francis}
}

@inproceedings{madampe:2024,
  title={The struggle is real! The agony of recruiting participants for empirical software engineering studies},
  author={Madampe, Kashumi and others},
  booktitle={2024 IEEE Symposium on Visual Languages and Human-Centric Computing (VL/HCC)},
  pages={417--422},
  year={2024},
  organization={IEEE}
}

@article{masood:2020,
  title={Real world scrum a grounded theory of variations in practice},
  author={Masood, Zainab and Hoda, Rashina and Blincoe, Kelly},
  journal={IEEE Transactions on Software Engineering},
  volume={48},
  number={5},
  pages={1579--1591},
  year={2020},
  publisher={IEEE}
}

@article{lenberg:2024,
  title={Qualitative software engineering research: Reflections and guidelines},
  author={Lenberg, Per and others},
  journal={Journal of Software: Evolution and Process},
  volume={36},
  number={6},
  pages={e2607},
  year={2024},
  publisher={Wiley Online Library}
}

@misc{rampton:2026,
  author       = {Rampton, John},
  title        = {Are Job Titles Losing Their Meaning? How Job Title Inflation Could Be Hurting Your Company},
  year         = {2026},
  month        = apr,
  url          = {https://www.linkedin.com/pulse/job-titles-losing-meaning-how-job-title-inflation-could-john-rampton-r7mac/},
  note         = {LinkedIn Pulse post}
}

@article{pant:2024,
  title={Ethics in the age of {AI}: An analysis of {AI} practitioners’ awareness and challenges},
  author={Pant, Aastha and Hoda, Rashina and Spiegler, Simone V and Tantithamthavorn, Chakkrit and Turhan, Burak},
  journal={ACM Transactions on Software Engineering and Methodology},
  volume={33},
  number={3},
  pages={1--35},
  year={2024},
  publisher={ACM New York, NY, USA}
}

@article{ferino:2026,
  title={Novice developers’ perspectives on adopting {LLMs} for software development: A systematic literature review},
  author={Ferino, Samuel and Hoda, Rashina and Grundy, John and Treude, Christoph},
  journal={ACM Transactions on Software Engineering and Methodology},
  year={2026},
  publisher={ACM New York, NY}
}

@article{gunatilake:2025,
  title={Manifestations of empathy in software engineering: How, why, and when it matters},
  author={Gunatilake, Hashini and Grundy, John and Hoda, Rashina and Mueller, Ingo},
  journal={IEEE Transactions on Software Engineering},
  year={2025},
  publisher={IEEE}
}

@article{madugalla:2024,
  title={Challenges, adaptations, and fringe benefits of conducting software engineering research with human participants during the covid-19 pandemic},
  author={Madugalla, Anuradha and Kanij, Tanjila and Hoda, Rashina and Hidellaarachchi, Dulaji and Pant, Aastha and Ferdousi, Samia and Grundy, John},
  journal={Empirical Software Engineering},
  volume={29},
  number={4},
  pages={86},
  year={2024},
  publisher={Springer}
}

@article{sindermann:2021,
  title={Assessing the attitude towards artificial intelligence: Introduction of a short measure in German, Chinese, and English language},
  author={Sindermann, Cornelia and others},
  journal={KI-K{\"u}nstliche intelligenz},
  volume={35},
  number={1},
  pages={109--118},
  year={2021},
  publisher={Springer}
}

@article{feng:2025,
  title={From Gains to Strains: Modeling Developer Burnout with GenAI Adoption},
  author={Feng, Zixuan and Afroz, Sadia and Sarma, Anita},
  journal={arXiv preprint arXiv:2510.07435},
  year={2025}
}

@article{sun:2025,
  title={Quality Assurance of LLM-generated Code: Addressing Non-Functional Quality Characteristics},
  author={Sun, Xin and St{\aa}hl, Daniel and Sandahl, Kristian and Kessler, Christoph},
  journal={arXiv preprint arXiv:2511.10271},
  year={2025}
}

@article{argerich:2024,
  title={Measuring and improving the energy efficiency of large language models inference},
  author={Argerich, Mauricio Fadel and Pati{\~n}o-Mart{\'\i}nez, Marta},
  journal={IEEE Access},
  volume={12},
  pages={80194--80207},
  year={2024},
  publisher={IEEE}
}

@article{hou:2024,
author = {Hou, Xinyi and others},
title = {Large Language Models for Software Engineering: A Systematic Literature Review},
year = {2024},
publisher = {Association for Computing Machinery},
address = {New York, NY, USA},
issn = {1049-331X},
url = {https://doi.org/10.1145/3695988},
doi = {10.1145/3695988},
journal = {ACM Trans. Softw. Eng. Methodol.},
month = sep
}

@ARTICLE{wang:2024,
  author={Wang, Junjie and others},
  journal={IEEE Transactions on Software Engineering}, 
  title={Software Testing With Large Language Models: Survey, Landscape, and Vision}, 
  year={2024},
  volume={50},
  number={4},
  pages={911-936},
  doi={10.1109/TSE.2024.3368208}}

@inproceedings{lee:2025,
  title={The Impact of Generative {AI} on Critical Thinking: Self-Reported Reductions in Cognitive Effort and Confidence Effects From a Survey of Knowledge Workers},
  author={Lee, Hao-Ping Hank and others},
  year={2025},
  booktitle={Proceedings of the CHI Conference on Human Factors in Computing Systems},
}

@article{di:2025,
  title={On the use of large language models in model-driven engineering},
  author={Di Rocco, Juri and others},
  journal={Software and Systems Modeling},
  pages={1--26},
  year={2025},
  publisher={Springer}
}

@inproceedings{alami2:2025,
  title={Human and Machine: How Software Engineers Perceive and Engage with {AI}-Assisted Code Reviews Compared to Their Peers},
  author={Alami, Adam and Ernst, Neil},
  booktitle={2025 IEEE/ACM 18th International Conference on Cooperative and Human Aspects of Software Engineering (CHASE)},
  pages={63--74},
  year={2025},
  organization={IEEE}
}

@article{kalliamvakou:2017,
  title={What makes a great manager of software engineers?},
  author={Kalliamvakou, Eirini and others},
  journal={IEEE Transactions on Software Engineering},
  volume={45},
  number={1},
  pages={87--106},
  year={2017},
  publisher={IEEE}
}

@misc{researchArtifact,
  author={Ferino, Samuel and Hoda, Rashina and Grundy, John and Treude, Christoph},
  title = { {Supplementary Information Package - STGT}},
  year = 2025,
  url = {https://doi.org/10.5281/zenodo.17556044},
  urldate = {2025-11-08}
}

@online{databricksgpt5:2025,
  author    = {Dogra, Archika and Nieto, Ana},
  title     = {Build with GPT-5 on Databricks with {AI} Gateway},
  year      = {2025},
  url       = {https://www.databricks.com/blog/build-gpt-5-databricks-ai-gateway},
  note      = {Accessed: 2025-10-07}
}

@online{mscopilot:2025,
  author    = {Microsoft team},
  title     = {Overview of Copilot for Power BI},
  year      = {2025},
  url       = {https://learn.microsoft.com/en-us/power-bi/create-reports/copilot-introduction},
  note      = {Accessed: 2025-10-07}
}

@article{jegham:2025,
  title={How hungry is {AI}? benchmarking energy, water, and carbon footprint of llm inference},
  author={Jegham, Nidhal and Abdelatti, Marwan and Elmoubarki, Lassad and Hendawi, Abdeltawab},
  journal={arXiv preprint arXiv:2505.09598},
  year={2025}
}

@article{tulili:2023,
  title={Burnout in software engineering: A systematic mapping study},
  author={Tulili, Tien Rahayu and Capiluppi, Andrea and Rastogi, Ayushi},
  journal={Information and Software Technology},
  volume={155},
  pages={107116},
  year={2023},
  publisher={Elsevier}
}

@article{burtch:2024,
  title={The consequences of generative {AI} for online knowledge communities},
  author={Burtch, Gordon and others},
  journal={Scientific Reports},
  volume={14},
  number={1},
  pages={10413},
  year={2024},
  publisher={Nature Publishing Group UK London}
}

@article{ozkaya:2023,
  title={Application of large language models to software engineering tasks: Opportunities, risks, and implications},
  author={Ozkaya, Ipek},
  journal={IEEE Software},
  volume={40},
  number={3},
  pages={4--8},
  year={2023},
  publisher={IEEE}
}

@inproceedings{mathews:2024,
author = {Mathews, Noble Saji and Nagappan, Meiyappan},
title = {Test-Driven Development and LLM-based Code Generation},
year = {2024},
isbn = {9798400712487},
publisher = {Association for Computing Machinery},
address = {New York, NY, USA},
url = {https://doi.org/10.1145/3691620.3695527},
doi = {10.1145/3691620.3695527},
booktitle = {Proceedings of the 39th IEEE/ACM International Conference on Automated Software Engineering},
pages = {1583–1594},
numpages = {12},
location = {Sacramento, CA, USA},
series = {ASE '24}
}

@article{chen:2025,
  title={An empirical study on challenges for llm application developers},
  author={Chen, Xiang and others},
  journal={ACM Transactions on Software Engineering and Methodology},
  year={2025},
  publisher={ACM New York, NY}
}

@article{shi:2025,
  title={Efficient and Green Large Language Models for Software Engineering: Literature Review, Vision, and the Road Ahead},
  author={Shi, Jieke and Yang, Zhou and Lo, David},
  journal={ACM Transactions on Software Engineering and Methodology},
  volume={34},
  number={5},
  pages={1--22},
  year={2025},
  publisher={ACM New York, NY}
}

@article{zhou:2024,
  title={Exploring the problems, their causes and solutions of {AI} pair programming: A study on GitHub and stack overflow},
  author={Zhou, Xiyu and others},
  journal={Journal of Systems and Software},
  pages={112204},
  year={2024},
  publisher={Elsevier}
}

@inproceedings{kabir:2024,
  title={Is stack overflow obsolete? an empirical study of the characteristics of chatgpt answers to stack overflow questions},
  author={Kabir, Samia and Udo-Imeh, David N and Kou, Bonan and Zhang, Tianyi},
  booktitle={Proceedings of the CHI Conference on Human Factors in Computing Systems},
  pages={1--17},
  year={2024}
}

@article{dakhel:2023,
  title={Github copilot {AI} pair programmer: Asset or liability?},
  author={Dakhel, Arghavan Moradi and others},
  journal={Journal of Systems and Software},
  volume={203},
  pages={111734},
  year={2023},
  publisher={Elsevier}
}

@article{dwivedi:2023,
  title={Opinion Paper:“So what if ChatGPT wrote it?” Multidisciplinary perspectives on opportunities, challenges and implications of generative conversational {AI} for research, practice and policy},
  author={Dwivedi, Yogesh K and others},
  journal={International Journal of Information Management},
  volume={71},
  pages={102642},
  year={2023},
  publisher={Elsevier}
}

@article{teubner:2023,
  title={Welcome to the era of chatgpt et al. the prospects of large language models},
  author={Teubner, Timm and others},
  journal={Business \& Information Systems Engineering},
  volume={65},
  number={2},
  pages={95--101},
  year={2023},
  publisher={Springer}
}

@online{quotient:2026,
  author = {{Quotient}},
  title = {{An AI Maturity Model for Engineering Teams}: A research-driven framework for understanding how engineering organizations adopt {AI}},
  year = {2026},
  url = {https://www.getquotient.com/ai-maturity-model},
  urldate = {2026-05-17}
}

@book{roller:2015,
  title={Applied qualitative research design: A total quality framework approach},
  author={Roller, Margaret R and Lavrakas, Paul J},
  year={2015},
  publisher={Guilford Publications}
}

\end{document}